\documentclass[aps,prb,reprint,superscriptaddress]{revtex4-1}
\usepackage{braket}
\usepackage{graphicx}
\usepackage{amsmath}
\usepackage{amssymb}
\usepackage{bbold}
\usepackage{bm}
\usepackage[usenames]{color}
\usepackage{tabularx}

\renewcommand{\vec}[1]{\mathbf{#1}}

\begin{document}


\title{
 Extrinsic Spin-Charge Coupling in Diffusive Superconducting Systems}
\author{Chunli Huang}
\affiliation{Department of Physics, The University of Texas at Austin, Austin, Texas 78712,USA}

\author{Ilya V.~Tokatly}
\affiliation{Nano-Bio Spectroscopy group, Departamento de Fsica de Materiales, Universidad del Pas Vasco, Av. Tolosa 72, E-20018 San Sebastin, Spain}
\affiliation{IKERBASQUE, Basque Foundation for Science, E-48011 Bilbao, Spain}
\affiliation{Donostia International Physics Center (DIPC),
Manuel de Lardizabal 4, E-20018 San Sebastian, Spain}

\author{F.~Sebastian Bergeret}
\affiliation{Donostia International Physics Center (DIPC),
Manuel de Lardizabal 4, E-20018 San Sebastian, Spain},
\affiliation{Centro de Física de Materiales (CFM-MPC), Centro Mixto CSIC-UPV/EHU,
Manuel de Lardizabal 5, E-20018 San Sebastián, Spain}

\date{\today}
\begin{abstract}
	We  present a theoretical study  of  diffusive superconducting systems with extrinsic spin-orbit coupling and arbitrarily strong impurity potential. We derive from a microscopic Hamiltonian a diffusion equation for the quasi-classical Green function, and  demonstrate that all  mechanisms related to the spin-orbit coupling are expressed in terms of three  kinetic coefficients: the spin Hall angle, the spin current  swapping coefficient, and the spin relaxation rate due to Elliott-Yafet mechanism.   The derived diffusion equation contains 
a hitherto unknown term describing a spin-orbit torque that appears exclusively in the superconducting state.
	As an example,  we provide a qualitative  description of  a magnetic vortex in a superconductor with triplet correlations, and show that the novel term describes a spin torque proportional to the vector product between the spectral angular momentum of the condensate and the triplet vector.  
	Our equation  opens up the possibility to explore spintronic effects in superconductors with no counterparts in the normal metallic state. 
\end{abstract}	
\maketitle

\section{Introduction}

Exotic phenomena can occur when two or more materials with different properties are merged into a single hybrid-structure.
Of particular interest are hybrid-structures consisting of conventional superconductor (S) and ferromagnet (F)
in which the interplay between these two quantum states leads to 
striking effects due to the appearance of odd-triplet superconducting correlations. \cite{RevModPhys.77.1321,buzdin2005proximity}.

The quasiclassical kinetic equation provides a unified description
of realistic hybrid-structures, including disorder, interfaces and out-of-equilibrium situations. This approach becomes particular relevant when the hybrid-structure is in the diffusive limit. The resulting diffusion equation describing superconducting structures is  known as the Usadel equation\cite{usadel1970generalized}. Its  extraordinary descriptive power  has been demonstrated by the agreements between theory and  experiments on S-N (N is a normal metal)\cite{belzig1999quasiclassical,le2008phase,dubos2001josephson} and S-F structures\cite{ryazanov2001coupling,bergeret2004induced,xia2009inverse}.  As such, the Usadel equation has been extended to discuss spin-relaxation induced by
spin-orbit coupling  (SOC) and/or magnetic exchange field. \cite{alexander1985theory}.
 In fact, since the discovery of the Knight shift (paramagnetic response of superconductor), spin-relaxation induced by SOC has been  extensively investigated in various superconducting systems \cite{PhysRevLett.3.325,soi_AG,fulde-maki1966,marsiglio2008superconductivity,PhysRevB.73.064505,PhysRevB.63.052501}.
Besides spin-relaxation \cite{glazov2010two},  SOC can also lead to fascinating spin-charge coupling phenomena that has spurred intensive research activities  in the field of spintronics.
 Notable spin-charge coupling phenomena like  the anomalous Hall effect \cite{Nagaosa_RevModPhys.82.1539},  the spin Hall effect \cite{sinova2015spin} (SHE) and the spin-galvanic effect \cite{IvcLyaPik1989,IvcLyaPik1990,Edelstein1990233,pikus1991spin} have been  discussed extensively in metallic systems.

Since the discovery of anomalous Hall effect, the microscopic origin of spin-charge coupling has been receiving everlasting attentions, see Ref.~\onlinecite{Nagaosa_RevModPhys.82.1539} for a historical overview. It is customary to separate the microscopic origin of SOC into the intrinsic and extrinsic type. While intrinsic SOC originates from inversion breaking potential that respects translational symmetry of the underlying material, the extrinsic SOC is generated from disorder potential that breaks translational symmetry such as random impurities.

 In a superconducting state, the effect of spin-charge coupling stemming from intrinsic SOC has been investigated extensively. For example, Ref.~\onlinecite{PhysRevLett.75.2004,PhysRevB.72.172501} discussed the Edelstein effect in non-centrosymmetric superconductors while Ref.~\onlinecite{PhysRevB.55.15174,PhysRevB.73.064505,Bobkova2004} discussed the consequences of uniform SOC in superconducting junctions. Two of the present authors also study the effect of intrinsic SOC in S/F hybrid structures \cite{bergeret2013,bergeret2014, tokatly2017usadel} and found that spin-charge coupling is related to the coupling between the singlet and triplet components of the superconducting condensate.  
%
%
%

Contrary to intrinsic SOC,  spin-charge conversion induced by extrinsic SOC received considerably lesser attention despite its importance in understanding superconducting junctions where disorder is ubiquitous.
Recently, some  quasiclassical kinetic theories have been  put forward to discuss the effect of spin-charge conversion generated by extrinsic SOC\cite{bergeret2016,brataas2017} . In Ref.~\onlinecite{bergeret2016}, a quasiclassical kinetic equation is developed using the Born approximation to treat the disorder induced self-energy. Since the Born approximation does not capture skew-scattering induced by SOC impurities \cite{Nagaosa_RevModPhys.82.1539}, the derived equation in Ref.~\onlinecite{bergeret2016} is parameterized by spin Hall angle $\theta$ that only contains contributions from the so-called side-jump mechanism.
In Ref.~\onlinecite{brataas2017}, the skew-scattering is taken into account by evaluating the self-energy at the third Born approximation but the side-jump mechanism is not fully accounted for and the resulting Usadel equation missed a spin torque generated by SOC (i.e.~Eq.~\eqref{eq:torque}) which, as we show below, is present in the superconducting state. 
 
In this work, we present a comprehensive and rigorous study of spin-charge conversion in superconductors with extrinsic SOC in the diffusive limit beyond the Born approximation. Our main result generalized the Usadel equation to account for the spin Hall and the spin current swapping effects. 
Strikingly, we uncover a non-linear and non-local torque induced by SOC that vanishes in the normal metallic state. 
 We demonstrate that all the terms related to spin-charge coupling in the Usadel equation can be parametrized by the spin-relaxation time and two kinetic coefficients -- the normal state spin Hall angle $\theta$ and the spin swapping \cite{lifshitz2009} coefficient $\kappa$.
We assume the extrinsic spin-orbit coupling to be weak but treat the impurity scalar scattering exactly without resorting to any finite order Born approximation. In other words, the disorder potential can be arbitrarily large with the (s-wave) scattering length taking any value from minus to plus infinity. On the one hand, this provides  a unified picture of extrinsic spin-charge coupling where the comparison between side-jump and skew-scattering is unambiguous. 
On the other hand, our approach captures resonant scattering induced by SOC disorder  \cite{Fert_resonant1,Fert_resonant2,Fert_resonant3}, which may  lead to an enhancment of spin-charge conversion, as in the case of graphene decorated with adatoms \cite{ferreira2014extrinsic,chunli2016}, or materials with Kondo impurities\cite{seki2008giant,Nagaosa09}. 


 The article is organized as  follow: In Sec.~\ref{sec:result}, we present and discuss our main result -- a generalization of the Usadel equation which accounts for extrinsic spin-charge coupling. In Sec.~\ref{sec:model},  we introduce our microscopic model, basis for the Green function, and the general formulation of  kinetic theory. In Sec.~\ref{sec:coll-int}, we describe the evaluation of the  collision integral in the presence of SOC and generalize the Usadel equation. Emphasis is placed on careful analysis of spatially non-local self-energies and the power-counting scheme to correctly capture the side-jump effect and the spin-orbit torque. 
 In Sec.~\ref{sec:discussion}, we illustrate the effect of spin-orbit torque in the presence of a vortex. We close the article with a outlook in Sec.~\ref{sec:summary}. Technical details concerning the evaluation of self-energy and collision integral are given in the Appendix. Throughout this work, we set $\hbar=1$ and adopt the rule of summing over repeated indices.

\section{Main Results}\label{sec:result}
%

We aim to derive a set of quasiclassical  kinetic equations for diffusive superconducting systems with extrinsic spin-orbit coupling. These equations are valid in the regime $\xi \gg l \gg p_{F}^{-1} \gg \lambda$ where $\xi$, $l$, $p_F^{-1}$, and $\lambda$ correspond to the superconducting coherence length, mean-free path, Fermi wavelength and the effective Compton wavelength of a material,  respectively. 

The quasiclassical approximation focus on the spatial variation of   observables and spectral functions over distances much larger than the Fermi wavelength $p_{F}^{-1}$. The diffusive limit further sets $\xi \gg l$. The SOC, being a relativistic phenomena, is microscopically associated with a material dependent effective Compton wavelength $\lambda$. The effective Compton wavelength can be much larger than its vacuum value \cite{Vignale2009}, but it is still significantly smaller than the Fermi momentum in most materials \cite{TakMae2008}, i.e.~$ \lambda^2 p_F^{2}  \ll 1$.Note the extrinsic SOC that leads to Mott scattering starts at power $\lambda^2 p_F^{2}$, this is different from an uniform SOC in electron gas where the expansion can occur at $\alpha p_F$ where $\alpha$ is the strength of uniform SOC. This justifies a perturbation theory in $\lambda^2 p_F^2$ while allowing the impurity potential strength to be arbitrarily large. In other words,  we 
sum the entire Born series generated by the impurity potential (see  Fig.~\ref{fig:sigma0}c). This introduces the \textit{scattering length} $a$ which, within our approach, can take any value $ -\infty<a<\infty$. 

Our main result is the generalized Usadel equation describing the effect of extrinsic SOC:
\begin{equation} \label{eq:Usadel}
\tau_{3}\partial_{t_1}\check{g}+ \partial_{t_2}\check{g} \tau_{3} +i [\check{\Delta} ,\check{g} ]+\partial_{k}\check{\cal{J}}_k
=  \frac{-1}{8\tau_{\mathrm{so}}} \left[\sigma^a \check{g} \sigma^a ,  \check{g} \right]+ \check{\cal{T}}.
\end{equation}
Here $\check{g}=\check{g}(\mathbf{r}, t_1,t_2)$  quasiclassical Green function (GF) averaged over momentum at the Fermi surface.  It is an $8\times 8$ matrix in the Nambu-spin-Keldysh space. The Pauli matrices $\sigma^a$ and $\tau_i$ span, respectively, the spin and Nambu spaces,  $\check{\Delta}$ is the anomalous superconducting self energy and $\tau_{\mathrm{so}}$ is the usual spin relaxation time induced by SOC disorder. 

\begin{figure}[t]
\centering
\includegraphics[width=0.5\textwidth]{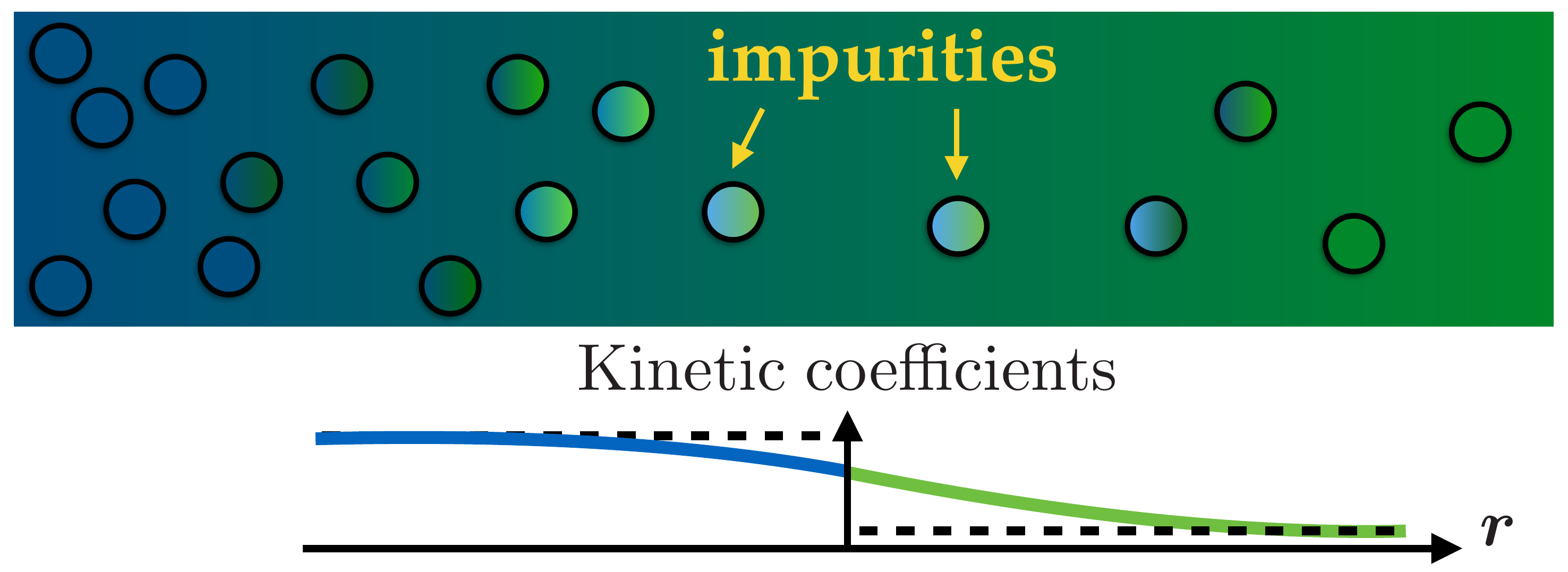}
\caption{The formulated kinetic theory (Eq.~\ref{eq:Usadel}) describes the diffusive transport of spin, charge and spectral weight in a superconducting hybrid structures. Importantly, our theory captures spatially varying kinetic coefficients (e.g.~spin-Hall angle $\theta$ and swap-current coefficient $\kappa$) which are important for spin-charge conversion. }
\label{fig:schematic}
\end{figure}

Nontrivial effects related to the  SOC enter the Usadel equation via the generalized $8\times 8$ matrix current $\check{\cal{J}}_k$ and the matrix torque $\check{\cal{T}}$. 
The matrix current $\check{\cal{J}}_k$ flowing in the spatial direction $k$ is defined as
\begin{equation} \label{eq:Jk}
\check{\cal{J}}_k = -D\left(\check{g} \partial_{k}\check{g} - \frac{\theta}{2} \epsilon_{akj}\big\{\partial_{j}\check{g},\sigma^{a}\big\} -i \frac{\kappa}{2} \epsilon_{akj}\left[\check{g}\partial_{j}\check{g},\sigma^{a}\right]\right)\; , 
\end{equation}
where $\epsilon_{kja}$ is the total antisymmetric tensor. {The first term in the right hand side of Eq. (\ref{eq:Jk}) is the standard diffusive current characterized by the diffusion constant $D=v_F\,l/3$.  The second  and third terms  describe the spin Hall effect (the anticommutator term couples  the charge and spin degrees of freedom) and the spin current swapping (the commutator couples different components of the spin flow), respectively. Note that our theory captures the slow variation of \emph{all} kinetic coefficients, $D$, $\theta$, $\kappa$ and $\tau_{\mathrm{so}}$ on scales larger than $p_F^{-1}$, as sketched in Fig. \ref{fig:schematic}.   The conservation of the generalized current $\check{\cal{J}}_k$ at interfaces between different materials define a boundary condition for Eq. (\ref{eq:Usadel})}.

Interestingly, we uncover a non-local spin-orbit torque $\check{\cal{T}}$ in  Eq.~(\ref{eq:Usadel}) when we account for the anomalous velocity induced by SOC disorder
\footnote{ There are  two typos in Ref.~\onlinecite{bergeret2016}:    First, in Eq. (5)  of that paper  the second term of our Eq.~\ref{eq:torque} was missing. Secondly, a factor $\lambda$ is missing in the definition of $\kappa$}. It is given by the following:
\begin{align} \label{eq:torque}
\check{\cal{T}}&=\frac{D}{4}\theta\,\epsilon_{akj}\,\big[\sigma^{a},\check{g}\partial_{k}\check{g}\partial_{j}\check{g}\big]+\frac{D}{4}\kappa\,\epsilon_{akj}\,i\big[\partial_{k}\check{g}\partial_{j}\check{g},\sigma^{a}].
\end{align}
While $\check{\cal{J}}_k$
, after taking corresponding traces,  
describes   charge and spin currents in the normal metallic state, $\check{\cal{T}}$   is only finite in the superconducting state  where the anomalous components of the GF are non-vanishing. Note that the trace of $\check{\cal{T}}$ over the spin Pauli matrices $\sigma$ are always zero.  It gives a finite contribution if  one  first multiplies  Eq. (\ref{eq:Usadel}) by  $\tau_3\sigma^a$ and then takes the trace, i..e.~$\check{\cal{T}}$ describes a novel type of spin torque in superconductors. 

Both the torque $\check{\cal{T}}$ and the generalized current  $\check{\cal{J}}_k$ are parameterized by the spin Hall angle $\theta$ and the swapping coefficient $\kappa$ derived in section \ref{sec:coll-int}. They are given by following expressions:
\begin{align}  \label{eq:thetamain}
\theta  & = \frac{2}{3}\omega_2\tau + 2\frac{\omega_1 \tau}{p_F \, l} ,  \\
\kappa  & =\frac{2}{3}\omega_1\tau -2\frac{\omega_2 \tau}{p_F \, l}  \label{eq:kappamain}\;,
\end{align}
where $\tau$ is the elastic scattering time. The effective spin-charge coupling rates
$\omega_1$ and $\omega_2$ can be expressed in terms of components of the single impurity scattering matrix at the Fermi energy: $ \hat{t}_{{\bf p}{\bf p}'}=A + i({\bf p}\times{\bf p}')\cdot \bm\sigma B/p_F^2$. Namely, $\omega_1 = 2\pi n_{\mathrm{im}} N_F \mathrm{Re} \big[ A^* B\big]$ and  $\omega_2= 2\pi n_{\mathrm{im}} N_F \mathrm{Im}\big[A^* B\big]$, where $N_F$ is the density of states at the Fermi energy, and $n_{\mathrm{im}}$ is the impurity concentration. To the lowest order in SOC, the coefficient $B$ is real, so that $\omega_2 $ is related to $\omega_1 $ via the optical theorem $\omega_2=p_F a \,\omega_1$, where $a$ is the scattering length.

Importantly, the kinetic coefficients $\theta$ and $\kappa$ are exactly those characterizing the coupled spin-charge transport in the normal state. 
The first and the second terms in Eq.~(\ref{eq:thetamain}) are the renowned skew scattering and the side-jump contributions to the spin Hall angle respectively. 
In Eq.~\eqref{eq:kappamain}, the first term was  identified  by Lifshitz and Dyakonov \cite{lifshitz2009} as the swap current coefficient, whereas the second term arises when we consistently include the anomalous velocity induced by SOC.  The latter  modifies the first term just as side-jump modifies  the  skew scattering in $\theta$.  

As it will become clear later, we shall name the first (second) term in Eq.~\eqref{eq:kappamain} as the $\textit{local}$ ($\textit{nonlocal}$) swap-current coefficient.  Similar to the side-jump contribution to the spin Hall conductivity, we found that the ``swap-current conductivity'' would also have a component that scales independently from the  impurity concentration. Note that in the limit of strong scattering potential $a \rightarrow \infty$, the skew-scattering  dominates over  the side-jump mechanism in $\theta$ while the nonlocal swap current  dominates the local mechanism in $\kappa$.

\section{Model Hamiltonian, basis and  kinetic formulation} \label{sec:model}
In this section, we discuss the model Hamiltonian,  the basis we use to define the  matrix Green functions, and the basic  kinetic theory of  Green function. The starting mean field Hamiltonian for a superconducting system  is $\int  d^3\mathbf{r} \,\mathcal{H}(\mathbf{r}) $, where
\begin{align}
\mathcal{H}(\mathbf{r})&=  \psi^{\dagger}_{\alpha}(\mathbf{r}) K_{\alpha \beta}(\mathbf{r},-i\partial_\mathbf{r}) \psi_{\beta}(\mathbf{r}) \nonumber \\
& +\frac{1}{2} \left( \psi^{\dagger}_{\alpha}(\mathbf{r})\Delta(\mathbf{r})i\sigma^2_{\alpha\beta} \psi^{\dagger}_{\beta}(\mathbf{r}) + \psi_{\alpha}(\mathbf{r})\Delta^{*}(\mathbf{r}) i\sigma^2_{\alpha\beta}\psi_{\beta}(\mathbf{r}) \right)\; .
\label{eq:Hamiltonian}
\end{align}
The mean-field superconducting order parameter $\Delta$ is local in space, $\sigma^2$ is the second Pauli matrix and $K_{\alpha \beta}(\mathbf{r},-i\partial_\mathbf{r})$ is the single-particle Hamiltonian given by  
\begin{equation}
K_{\alpha\beta}(\mathbf{r},-i\partial_\mathbf{r})= \left[-\frac{\nabla^2}{2m}-\mu \right]\delta_{\alpha\beta} + \mathcal{V}_{\alpha \beta}(\mathbf{r},i\partial_\mathbf{r}).
\end{equation}
We consider here a disorder potential $\mathcal{V}_{\alpha\beta}(\mathbf{r},i\partial_\mathbf{r})$ which contains a spin-independent part, proportional to $\delta_{\alpha\beta}$,  and a spin-orbit coupling part porportional to $\sigma_{\alpha\beta}$:
\begin{align} \label{eq:imp_model}
\mathcal{V}_{\alpha\beta}(\mathbf{r},i\partial_\mathbf{r})=    U(\mathbf{r}) \delta_{\alpha\beta}+ \lambda^2\sigma_{\alpha\beta}\cdot (\nabla U(\mathbf{r})  \times -i\nabla  )
\end{align}
\begin{equation}
 U(\mathbf{r})= \sum_{a}^{N_i} V(\mathbf{r}-\mathbf{r}_a)
 \label{scal_imp}
\end{equation}
Here $\lambda$ is the material dependent Compton wavelength and we assume $\lambda^2 p_F^2 \ll 1$. $V(\mathbf{r})$ is a short-range potential induced by $N_i$ randomly distributed impurities. 
%

The   field operators entering Eq. (\ref{eq:Hamiltonian}) can be conveniently organized as a spinor: 
\begin{equation}
\Psi=\left(
\begin{array}{c}
\psi_\uparrow\\
\psi_\downarrow\\
\end{array}
\right)\label{spinor}
\end{equation}
and correspondingly 
\begin{equation}
\Psi^\dagger=\left(
\begin{array}{c}
\psi^\dagger_\uparrow\\
\psi^\dagger_\downarrow\\
\end{array}\label{spinor2}
\right)
\end{equation}
It is  customary, see for example  Ref.~\onlinecite{alexander1985theory} or the chapter by K. Maki in Ref.~\onlinecite{parks1969superconductivity}, to define  the matrix Green functions as the time ordered correlator  $\check G=-i\langle T \mathbf{\Psi}\mathbf{\Psi}^\dagger\rangle$ of the product between the column bi-spinor  $\mathbf{\Psi}=(\psi_\uparrow,\psi_\downarrow,\psi^\dagger_\uparrow,\psi^\dagger_\downarrow)$ and the row bispinor  $\mathbf{\Psi}^\dagger=(\psi_\uparrow^\dagger,\psi_\downarrow^\dagger,\psi_\uparrow,\psi_\downarrow)$.  The Gorkov equations  for this  GF can then be obtained straightforwardly by  using the Heisenberg equation of motion  of the field operators \cite{abrikosov1963methods}.

Although  the GF defined in the above  basis  are  widely used in the literature, we opt here for another basis which allows for a more intuitive interpretation of different terms in the equation of motion for the GF and subsequently in the kinetic equation.  Instead of using the spinors defined in Eq.~(\ref{spinor}) and (\ref{spinor2}) we introduce the  time-reversal conjugated spinor
\begin{equation}
\Psi^c=i\sigma^y\Psi^\dagger=
\left(
\begin{array}{c}
\psi^\dagger_\downarrow\\
-\psi^\dagger_\uparrow\\
\end{array}\label{spinor3}
\right)
\end{equation}
and construct the GF as $\check G=-i\langle T \tilde{\mathbf{\Psi}}\tilde{\mathbf{\Psi}}^\dagger\rangle$, where $\tilde{\mathbf{\Psi}}=(\Psi,\Psi^c)$ and 
$\tilde{\mathbf{\Psi}}^\dagger=(\Psi^\dagger,\Psi^{\dagger c})$. A more intuitive form of the equations of motion for such defined GF is related to the fact that it explicitly reflects the superconducting paring between the time-reversal conjugated states.
In our basis, the quasiclassical matrix GF that enters   Eq. (\ref{eq:Usadel}) has the form:
\begin{align} \label{eq:qc_GF}
\check g = \begin{pmatrix} \hat g& \hat f\\
-\hat f^c & -\hat g^c
\end{pmatrix},
\end{align}
where the hat $\hat{.}$ quantities are matrices in spin space:
\begin{eqnarray}
\hat g&=&g_0+g^a\sigma^a,\\
\hat f&=&f_0+f^a\sigma^a.
\end{eqnarray}
One of the advantages of using this basis is that the Pauli matrices only appear multiplying spin-related quantities, in particular, the triplet components  of the condensate amplitude $f^a$ where $a=x,y,z$.
 Furthermore, since the kinetic energy and the impurity potential is time-reversal invariant [i.e.~$K(\mathbf{r},-i\partial_\mathbf{r})=\sigma^{y}K(\mathbf{r},i\partial_\mathbf{r})\sigma^y$],  it is simply proportional to identity in Nambu space $\tau_0$, when written in our basis.
 
In contrast, if one uses the GF defined by the basis in Eq.~(\ref{spinor})-(\ref{spinor2}), all components of  the anomalous Green's function acquire an additional  $i\sigma_y$ factor. Moreover, all spin-dependent fields has to be written using a Nambu  diagonal matrix proportional to ${\rm diag}[\vec\sigma,\vec\sigma^*]$ \cite{alexander1985theory}. In Table I, we compare different physical quantities expressed in the basis used in  Ref.~\onlinecite{alexander1985theory} and our basis.
For readers who wish to recover the GF defined in Ref.~\onlinecite{alexander1985theory}, they can do so by applying the following transformation to our matrix GF defined in Eq.(\ref{eq:qc_GF}):  $\check U^{\dagger}\check g\check U$  with $U=\frac{1}{2}(1+i\sigma_{y})(1-i\tau_{3}\sigma_{y})=e^{i\frac{\pi}{4}\sigma^{y}(1-\tau_{3})}$. Correspondingly one can use this transformation to transform our  Usadel equation, Eq.~(\ref{eq:Usadel}),  to  the basis used  in Refs. \onlinecite{alexander1985theory,lofwander2005interplay,alidoust2010spin}.

Having established the basis in which  the Green functions are written, we now derive the kinetic equation governing the charge-spin coupling in superconducting systems. 
The derivation of the quasiclassical kinetic equation from microscopic Hamiltonian can be found in many textbooks \cite{kamenev2011field,rammer2007quantum} and reviews \cite{danielewicz1984quantum,rammer1986quantum,RevModPhys.77.1321}. 
Here,  we provide a brief summary of it and postpone the calculation of the collision integral within the
  quasiclassical approach to the next section. 
Given a microscopic Hamiltonian [Eq.~\eqref{eq:Hamiltonian}],  one derives the left and right Dyson equation of the GF using the Heisenberg equation of motion. The standard starting point to derive the kinetic equation is to consider the left-right substrated Dyson equation:
\begin{align} \label{eq:KE1}
&\tau_{3}\partial_{t_1}\check{G}(1,2)+  \partial_{t_2}\check{G}(1,2) \tau_{3}
+\frac{\left( \nabla_{1}^2 -\nabla_{2}^2 \right)}{2m}\check{G}(1,2) \nonumber  \\
&+i\check{\Delta}( \mathbf{r_1})\check{G}(1,2)-i\check{G}(1,2)\check{\Delta}( \mathbf{r_2}) \nonumber \\
&=-i \int d3 \,\, \check\Sigma(1,3)\check{G}(3,2)-\check{G}(1,3)\check\Sigma(3,2).
\end{align}
where for abreviation the numbers $j=1,2,3$ denote the set of spatial and time coordinates $\mathbf{r}_j,t_j$. Here $\check{\Sigma}$ is the self-energy.
In the  basis we have chosen to represent the Green's functions  the matrix  describing the superconducting order parameter 
reads
\begin{equation}
\check{\Delta}(\mathbf{r})= 
\begin{pmatrix}0 & \Delta(\mathbf{r})\\
-\Delta^{*}(\mathbf{r}) & 0
\end{pmatrix}.
\end{equation}
In order to derive the kinetic equation from Eq.~(\ref{eq:KE1}), one introduce the Wigner coordinates $\mathbf{r}=(\mathbf{r}_1+\mathbf{r}_2)/2$, 
$\mathbf{s}= \mathbf{r}_1-\mathbf{r}_2$. Unlike the derivation of the Boltzmann equation \cite{in-prep}, there is no obvious advantage of introducing the Wigner coordinates for the time component. Therefore, we  Fourier transform Eq.~(\ref{eq:KE1}) only with respect to the relative space coordinate $\mathbf{s}$ and arrive at the following equation:
%
%
\begin{widetext}
\begin{align} \label{eq:KE}
&\tau_{3}\partial_{t_1}\check{G}_\mathbf{p}(\mathbf{r})+ \partial_{t_2}\check{G}_\mathbf{p}(\mathbf{r}) \tau_{3} +\frac{p_i}{m} \partial_{i}\check{G}_\mathbf{p}(\mathbf{r})
+i [ \check{\Delta}(\mathbf{r}) ,   \check{G}_\mathbf{p}(\mathbf{r})] 
 \nonumber  \\
&= -i \left[ \check{\Sigma}_\mathbf{p}(\mathbf{r}) , \check{G}_\mathbf{p}(\mathbf{r}) \right] +   \left\{ \partial_j  \check{\Sigma}_\mathbf{p}(\mathbf{r}) ,  \partial_{p_j} \check{G}_\mathbf{p}(\mathbf{r})  \right\} 
-  \left\{ \partial_{p_j} \check{\Sigma}_\mathbf{p}(\mathbf{r}) ,  \partial_j \check{G}_\mathbf{p}(\mathbf{r})  \right\} \;.
\end{align}
\end{widetext}
Note that the (two) time arguments are skipped for brevity and we neglect the small spatial dependence of the gap function. More importantly, we 
 retain the Poission bracket in the right hand side that is usually discarded\cite{bergeret2007scattering}. It turns out that this term is essential to describe effects associated to charge-spin coupling \cite{bergeret2016}. Fundamentally, this is because the self-energy describing spin-charge coupling is not only made up by GFs that are local in space. As we shall see in the next section, the presence of SOC generates self-energy terms  that depends on nonlocal GFs. Hence, the commutator and the Poisson bracket on the right hand side of Eq.~\ref{eq:KE} can generate terms that are of the same order in the nonlocality of the GF.  

Eq.~\eqref{eq:KE} has the form of a  kinetic equation; it describes a balance between the driving force (left hand side) and collision integral (right hand side). Note that we have not given any prescription to perform the (quantum) average of the fermion field operators in the Green functions.  At zero-temperature, the average is taken over the ground state of a filled Fermi sea. At finite-temperature and thermal equilibrium, the average can be taken over the grand canonical ensemble using the Matsubara formalism. For a system that is out-of-equilibrium, we shall place the time-coordinates onto the Keldysh time contour and promote all 4$\times$4 matrices in Nambu-spin space onto an 8$\times$8 matrix in the Nambu-spin-Keldysh space. Since in all these situations Eq.~\eqref{eq:KE} remains formally unchanged,  we use the check symbol,  $\check .$, to denote either the  4$\times$4 matrices in the equilibrium case or  8$\times$8 matrices in the Keldysh formalism.

%
%



\section{Derivation of the Usadel equation} \label{sec:coll-int}

The anitcommutator in Eq.~\eqref{eq:KE} describes important spin-charge coupling also poses a hurdle to continue the derivation of the kinetic equation following standard approach  \cite{kamenev2011field,rammer2007quantum,danielewicz1984quantum,rammer1986quantum,RevModPhys.77.1321}. 
In this section, we discuss in detail how we deviate from the standard approach and derive the generalized Usadel equation in the presence of disorder SOC from Eq.~\eqref{eq:KE}.

Let us begin by reminding the readers that Eq.~\eqref{eq:KE} still contains superfluous information that is unessential for the description of electronic transport near the Fermi energy. In a superconductor, where the density of states  changes dramatically around the Fermi energy, it is customary to simplify Eq.~\eqref{eq:KE} within the quasiclassical approximation. 
In this  approximation, the Fermi-energy is the largest energy scale in the problem and, as mentioned in Sec. \ref{sec:result}, spatial variations of all observables and spectral functions take place over distances much larger than the inverse of the  Fermi momentum. 
 Moreover, the GFs are peaked at the Fermi level and therefore it is convenient to integrate them
 over the quasiparticle energy ($\xi_p$) to obtain the so-called quasiclassical Eilenberger Green function:
 \begin{equation} \label{eq:Eilen}
\check{\mathrm{g}}({\bf n}, \mathbf{r}) \equiv  \frac{i}{\pi}\int d\xi_p \, \check{G}_\mathbf{p}( \mathbf{r})\; ,
\end{equation}
where ${\bf n}$ is a unit vector pointing in the direction of the momentum at the Fermi surface.

The  standard way of deriving the quasiclassical kinetic equation is to integrate Eq.~\eqref{eq:KE} over the quasiparticle  energy  and to obtain an equation for $\check{\mathrm{g}}$, the Eilenberger equation\cite{larkin1969quasiclassical,alexander1985theory}. This  equation is complemented by a normalization condition $\check{\mathrm{g}}^2=1$.   In the present case however, the situation is more complicated and one cannot follow this path straightforwardly. 
This is because the Poisson bracket, {\it  i.e.}~the anti-commutators on the right hand side of Eq.~\eqref{eq:KE}, contains momentum derivatives. They prohibit a straightforward integration over the quasiparticle energy  and do not ensure the normalization condition for the GF at this stage.

In order to overcome these difficulties  we follow the procedure 
put forward  in Ref. \onlinecite{bergeret2016} and assume that the system is in the \textit{diffusive} regime. In this limit the system is almost isotropic in space. We then  expand $\check{g}$ in spherical harmonics and keep only the zeroth and first moments:
\begin{equation} \label{eq:ansatz}
\check{\mathrm{g}}(\mathbf{n}, \mathbf{r})\approx \check{g}( \mathbf{r})+ n_k \check{g}_{k}( \mathbf{r}).\; 
\end{equation}

Our goal is to obtain a close equation for the zeroth-moment Green function $\check{g}(\vec{r})$, {\it i.e}. the Usadel Green function. 
For this sake, we resort to the following counting scheme of small parameters in the diffusive limit. Let $\epsilon_0$, $\epsilon_1$ and $\epsilon_2$ describe the characteristic magnitudes of the zeroth moment, first moment and leading non-locality of the quasiclassical GF:
\begin{align} \label{eq:counting1}
  \check{g}(\mathbf{r})  &=  \mathcal{O} (\epsilon_0), \\
   \check{g}_k(\vec{r})  &= \mathcal{O}  (\epsilon_1),\\
  p_F^{-1} \partial_k  \check{g}(\vec{r}) &=\mathcal{O} (\epsilon_2). \label{eq:counting3}
\end{align}
In the diffusive limit, the zeroth moment of the GF is the  dominant component and we have the following hierarchy of scales $\epsilon_0 \gg \epsilon_1 \gg \epsilon_2 $. The first inequality $\epsilon_0 \gg \epsilon_1$ arises from the fact that we are considering the diffusive limit, $1 \gg l/\xi$,  while the second inequality,  $\epsilon_1 \gg \epsilon_2$, arises from the quasiclassical limit $  l,\xi\gg p_{F}^{-1}$. 
As mentioned in Sec.~\ref{sec:result}, our theory describes the macroscopic in-homogeneity of the disorder potential and we shall assume  that the kinetic coefficients changes on the scale of $\epsilon_2$.

 Spin-charge coupling occurs at linear order in   $\lambda^2 p_F^{2}$ 
 and has contributions  from both skew-scattering and side-jump mechanism.
The skew-scattering mechanism occurs at $\epsilon_1$; it does not require spatial nonlocality and can be captured in standard T-matrix calculation with equilibrium/uniform Green function. Unlike skew-scattering mechanism, the side-jump mechanism requires the Green function to be non-uniform in space so it is of the order  $\epsilon_1\epsilon_2$. 
Since in our power counting scheme $\epsilon_1 \gg \epsilon_2$,  in order to catch consistently the side-jump contribution, we also have to retain terms of order  $ \epsilon_{1}^{2}$.
These  terms  are typically discarded in the standard derivation of the Usadel equation without SOC.

At order $\lambda^4p_F^4$, the most dominant contribution to the self-energy is the Elliott-Yafet spin relaxation which occurs at order $\epsilon_0$. Hence, at this order, we shall only retain the $\epsilon_0$ term and neglect all other corrections arising from $\epsilon_1$ and $\epsilon_2$.

Having established our approximation scheme, we can proceed to compute the equation of motion for the zeroth and first moment from  Eq.~(\ref{eq:KE}). The resulting  equation of motion for $\check{g}(\vec{r})$ and $\check{g}_k(\vec{r})$ are given by:
\begin{align} 
\tau_{3}\partial_{t_1}\check{g} + \partial_{t_2}\check{g} \tau_{3} +\frac{v_F}{3} \partial_{k}\check{g}_k 
+i\left[\check{\Delta}  ,\check{g} \right]
&=\check{\mathcal{I}}_0 [\check{g} ,\check{g}_k], \label{eq:g_0}\\
 \frac{v_F}{3}  \partial_{k} \check{g} 
&=\check{\mathcal{I}}_{k} [\check{g} ,\check{g}_k] \label{eq:g_k}\; .
\end{align}
In order to lighten the notations, from now and what follows, we shall display the space arguments of $\check{g}$ and $\check{g}_k$ only when it is important for discussion.
  In Eq.~(\ref{eq:g_k}) we assume that the elastic scattering rate is much larger than the superconducting gap and the typical rate of change of $\check{g}_k$ i.e.~$\partial_{t_1},\partial_{t_2},\Delta \ll \tau^{-1}$.

The right hand side of  Eq.~\eqref{eq:g_0} and \eqref{eq:g_k} correspond to the collision integral of the zeroth and first moment GF respectively. They are the essence of extrinsic spin-orbit coupling:
\begin{align}  \label{eq:coll-inta}
\check{\mathcal{I}}_0 [\check{g} ,\check{g}_k] &=  -i \left\langle 
\left[\check\Sigma ,\,\check{\mathrm{g}}(\vec{n},\vec{r}) \right] \right\rangle  
 - \frac{\partial_{i}}{2} \left\langle  \big\{ \partial_{p_i} \check\Sigma \,, \, \check{\mathrm{g}}(\vec{n},\vec{r})\big\}      \right\rangle  \\
 \check{\mathcal{I}}_{k} [\check{g} ,\check{g}_k] &= -i \langle 
 n_k \left[\check \Sigma ,\,\check{\mathrm{g}}(\vec{n},\vec{r})  \right]\rangle \label{eq:coll-intb}
\end{align}
%
%
%
The GFs on the 
 right hand side of Eq.~\eqref{eq:coll-inta} and \eqref{eq:coll-intb} are the Eilenberger Green function, $\check{\mathrm{g}}(\vec{n},\vec{r})$. Their  arguments are shown explicit to distinguish them from the zeroth moment Green function $\check{g}$,  Eq.~(\ref{eq:ansatz}). The angular brackets in Eqs.~(\ref{eq:coll-inta})-(\ref{eq:coll-intb})  stand for integration over  the solid angle at the Fermi surface.
 In deriving Eq.~\eqref{eq:coll-inta}, we assume that  the Green function is zero at large momentum and used integration by parts to shift the momentum derivative from $\check{G}$ to $\check{\Sigma}$, {\it c.f.}~the second last term of Eq.~\eqref{eq:KE}. Within the quasiclassical approach, the momentum derivative on the self-energy is also evaluated on the Fermi surface, \begin{equation}
 \partial_{p_i} \check{\Sigma} =\partial_{p_i} \check{\Sigma}(\vec{p},\vec{r}) \big|_{p_i=p_F n_i}
 \end{equation}
Usually, the self-energy can be expressed as local GF ($\check{\Sigma}(\vec{p},\vec{r}) \propto \check{G}(\vec{p},\vec{r}) \,$), then the  second term in Eq.~\eqref{eq:coll-inta} resembles the familiar renormalization of the Fermi velocity due to self-energy $v_F \rightarrow v_F+ \partial_{p_i} \check{\Sigma}(\vec{p},\vec{r})$. However, the self-energy itself can also be a function of nonlocal GF ($\check{\Sigma}(\vec{p},\vec{r}) \propto\partial_i \check{G}(\vec{p},\vec{r}) \,$) in the presence of SOC disorder, so one has to account for the first term in Eq.~\eqref{eq:coll-inta} during the identification of the Fermi velocity renormalization, as described in Sec.~\ref{sec:Usadel}.
 In Eq.~\eqref{eq:coll-intb}, we neglect the Poisson bracket (i.e.~linear order in gradient) term. This is because, as we will show below, when the linear in SOC self-energy is substituted into the Poisson bracket, it generates terms of the order of $\epsilon_1^3$ and $\epsilon_1^2 \epsilon_2$, which are neglected in our approximation scheme.  This is not the case for the anticommutator  Eq.~\eqref{eq:coll-inta} which has to be kept.

 From Eq.~\eqref{eq:g_0} and \eqref{eq:g_k}, we derive the   the Usadel  equation as follows. First  we  evaluate the collision integrals  by expanding the  self-energy $\check{\Sigma}$  in terms of the small parameter  $\lambda^2 p_{F}^2$ up  to second order (see next subsections):
\begin{equation}
\check{\Sigma} = \check{\Sigma}^{(0)}+ \check{\Sigma}^{(1)} + \check{\Sigma}^{(2)},
\label{eq:Sigma_all}
\end{equation}
where $\check{\Sigma}^{(n)} \propto (\lambda p_F)^{2n}$. The zeroth order self-energy describes the usual Drude relaxation. The first and second order describes spin-charge coupling and the  Elliott-Yafet spin relaxation process respectively.  
Then, we substitute all the self-energies in Eq.~\eqref{eq:Sigma_all} into Eq.~\eqref{eq:coll-intb} to  express the first moment $\check{g}_k$ in terms of the zeroth moment $\check{g}$, and obtain the  so-called constitutive relation, Eq.~\eqref{eq:bare-current}.  From this  equation, we can infer the normalization condition $\check{g}^2=1$.
 Next, we substitute Eq.~\eqref{eq:Sigma_all} into Eq.~\eqref{eq:coll-inta} and identify the anomalous current $\check{\cal	J}_k^{\mathrm{an}}$ and the  spin-orbit torque $\check{\cal T}$. Lastly, we substitute the constitutive relation into Eq.~\eqref{eq:g_0} and arrive at the generalized Usadel equation.  
 
 We shall now  follow the procedure described above and evaluate various self-energies in  Eq.~(\ref{eq:Sigma_all}).

\subsection{Calculations of the self-energy $\check{\Sigma}$}

\subsubsection{$\check{\Sigma}^{(0)}$: Drude-relaxation \label{sec:sigma0}}

\begin{figure}[t]
\centering
\includegraphics[width=0.45\textwidth]{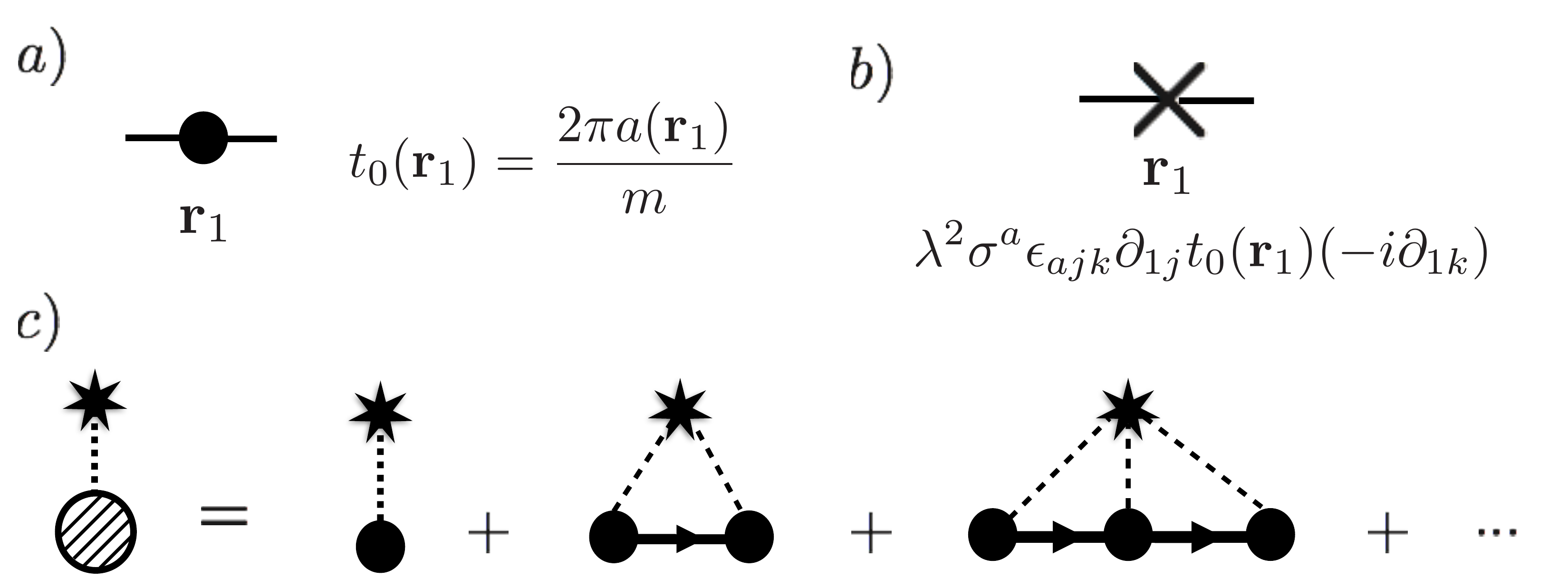}
\caption{ The scattering vertex of a) the spin-independent potential and b) the spin-orbit coupling potential acting on the annihilation field operator $ \tilde{\mathbf{\Psi}}$ in the quasiclassic formulation. Subplot c) depicts the self-energy $\check{\Sigma}^{(0)}$ in Eq.~\eqref{eq:Sigma-drude}. The star symbol represents the impurity density $n_{\mathrm{im}}$ while the arrow line represents  the  Green functions. This ``blob'' diagram is later used to compose the self-energy of spin-charge coupling, $\check{\Sigma}^{(1)}$ .  \label{fig:sigma0}
}
\end{figure}

In this section, we evaluate the self-energy induced by spin-independent scattering potential. Throughout this article, the concentration of  impurities, $n_{\mathrm{im}}$, is assumed to be small.  In this dilute impurity limit, the self-energy depends only on the scattering properties of a single impurity, i.e.~non-crossing approximation, see Fig.~\ref{fig:sigma0}. Let us first neglect  SOC and introduce the T-matrix describing the total scattering amplitude for an electron scattered by the scalar part of the impurity potential:
 \begin{align} \label{eq:T-matrixR}
\check T_{\vec{k} \vec{k'}}^{(0)}(\vec{r})=V_{\vec{k} \vec{k'}}+\sum_{\vec{p}} V_{\vec{k} \vec{p'}}\check G_{\vec{p}}(\vec{r}) \check T_{\vec{p} \vec{k'}}^{(0)}(\vec{r}) \; ,
 \end{align}
where $V_{\vec{k} \vec{k'}}=V({\bf k}-{\bf k}')$ is the Fourier component of the single impurity potential in Eq.~(\ref{scal_imp}), and the superscript $(0)$ reflects zeroth order in SOC. Here $\check G_{\vec{p}}(\vec{r}) $ is the GF in the Wigner representation that enters the full kinetic equation before quasiclassical approximation, Eq.~\eqref{eq:KE}.

In Eq.~\eqref{eq:T-matrixR}, we assume a short-range impurity potential with a dominating s-wave scattering. Under this assumption, one can use the standard renormalization procedure to eliminate the high energy contribution to the momentum integral in Eq.~(\ref{eq:T-matrixR}) by introducing the physical scattering amplitude $t_0=2\pi a/m$ at zero energy, where $a$ is the zero-energy scattering length. As a result, the T-matrix is expressed in terms of the  quasiclassical GF $\check g$ which describes quasiparticle dynamics in the vicinity of the Fermi surface, and the physical scattering amplitude $t_0$. The corresponding equation for $\check{T}^{(0)}$, which now describes the scattering of  quasiparticle at the  Fermi surface reads 
 \begin{equation}
 \check{T}^{(0)}( \mathbf{r})=t_{0}-i\pi N_F t_{0} \check{g}( \mathbf{r}) \, \check{T}^{(0)} ( \mathbf{r})\; .\label{eq:T_matrix_qc}
 \end{equation}
Here $N_F= m p_F /2\pi^2$ is the density of states at the Fermi energy. Note, since the dominant scattering wave induced by the impurity potential is assumed to be s-wave, only the isotropic part of the quasiclassical GF (the first term in Eq.~\ref{eq:ansatz}) enters the  $\check{T}^{(0)} $.
To leading order in impurity density $n_{\mathrm{im}}$, the self-energy can  be expressed in terms of $\check{T}^{(0)}$ as  follow ({\it cf.} Fig. \ref{fig:sigma0}c):
\begin{align} \label{eq:Sigma-drude}
\check{\Sigma}^{(0)}( \mathbf{r_1}, \mathbf{r_2})&=n_{\mathrm{im}} \delta( \mathbf{r_1}- \mathbf{r_2}) \check{T}^{(0)}( \mathbf{r_1}).
\end{align}

The quasiclassical GF $\check{g}$ is a $8\times 8$ matrix in spin-Nambu-Keldysh space, which, in the absence of SOC,  obeys the normalization condition $ \check{g}^2( \mathbf{r})=1$. Using this condition, one can solve Eq.~(\ref{eq:T_matrix_qc}) explicitly:
\begin{equation} \label{eq:fullT0}
\check{T}^{(0)} ( \mathbf{r}) = t_0\frac{1-ip_Fa\;\check{g}( \mathbf{r})}{1+(p_Fa)^2}= \mathrm{Re }\,t_F + i \,\mathrm{Im }\,t_F\check{g} ( \mathbf{r}) \; ,
\end{equation}
where we used the identity $\pi N_Ft_0=p_Fa$, and introduced the scattering amplitude $t_F$ of an electron at the Fermi energy in a normal equilibrium system:
\begin{equation}
\label{eq:opsa}
t_F= \frac{t_0}{1+ i\, p_F a}.
\end{equation}
%
%
%

It is worth  mentioning that since the above expressions are local in space (i.e.~they do not involve spatial derivatives), they are also valid for systems with spatially dependent concentration and/or type of impurities as schematically shown in Fig.~\ref{fig:schematic}. In other words, the impurity concentration $n_{\mathrm{im}}$,  the scattering amplitudes $t_{0}$ and $t_{F}$, and the scattering length $a$ may depend on the (slowly varying) spatial coordinate ${\bf r}$.

 Unlike $t_0( \mathbf{r})$, $t_F( \mathbf{r})$ is  a complex number and its complex phase satisfies the optical theorem:
\begin{equation} \label{eq:opti-theorem}
\mathrm{Im} \, t_F( \mathbf{r}) = -\pi N_F |t_F( \mathbf{r})|^2.
\end{equation}
After substituting Eq.~(\ref{eq:fullT0}) into Eq.~\eqref{eq:Sigma-drude} and performing the Fourier transform with respect to the difference of the coordinates we arrive at the (zeorth order in SOC) self-energy in the Wigner representation:
\begin{equation}  \label{eq:sigma_0}
\check{\Sigma}^{(0)}( \vec{n},\mathbf{r})=\check{\Sigma}^{(0)}( \mathbf{r})= n_{\mathrm{im}}  \mathrm{Re } \, t_F( \mathbf{r}) - \frac{i}{2\tau( \mathbf{r})} \,\check{g} ( \mathbf{r}).
\end{equation}
The elastic (Drude) relaxation time is expressed in terms of  $t_{F}( \mathbf{r})$ which can model arbitrarily strong impurity potential:
\begin{equation}
\frac{1}{\tau( \mathbf{r})} = 2\pi n_{\mathrm{im}}({\bf r})  N_F |t_F( \mathbf{r})|^2.
\end{equation}
In the next two sections  we analyze the SOC scattering at the impurities that leads to the spin-charge coupling and spin-relaxation.   We should point out that a  small contribution arising from SOC (i.e.~$\lambda^4 p_F^4$ or smaller) are  neglected from the Drude relaxation time.

\subsubsection{$\check{\Sigma}^{(1)}$: Spin-charge coupling} 
\label{sec:SCC}

Let us now include extrinsic SOC perturbatively.  To  leading order in SOC, we parametrize the renormalized spin-dependent part of the scattering vertex as 
\begin{align} \label{t-so}
t_{so}(\vec{r})=-i \lambda^2 \epsilon_{ajk}\sigma_{a} \partial_{j} t_0(\mathbf{r})  \partial_{k} \;,
\end{align}
which has the same form as the SOC term in Eq.~\eqref{eq:imp_model}, but with the bare impurity potential replaced with the zero-energy scattering amplitude $t_0$.  In particular, this means that the total scattering amplitude for electrons on the Fermi surface in the normal phase is approximated as 
$ \hat{t}_{{\bf p}{\bf p}'}= t_F + i \lambda^2 {\bm\sigma} \cdot ({\bf p}\times{\bf p}')t_0,
$
where $t_F$ is given by Eq.~(\ref{eq:opsa}). In other words, in the general form of the Mott scattering T-matrix $\hat{t}_{{\bf p}{\bf p}'}= A + i{\bm\sigma} \cdot  ({\bf p}\times{\bf p}')B/p_F^2$, the scalar part $A=t_F$ is the full complex scattering amplitude of a scalar potential, while the coefficient $B=\lambda^2p_F^2 t_0$ is purely real, which corresponds to the leading perturbative correction due to SOC. 

 \begin{figure}[t]
\includegraphics[width=0.45\textwidth]{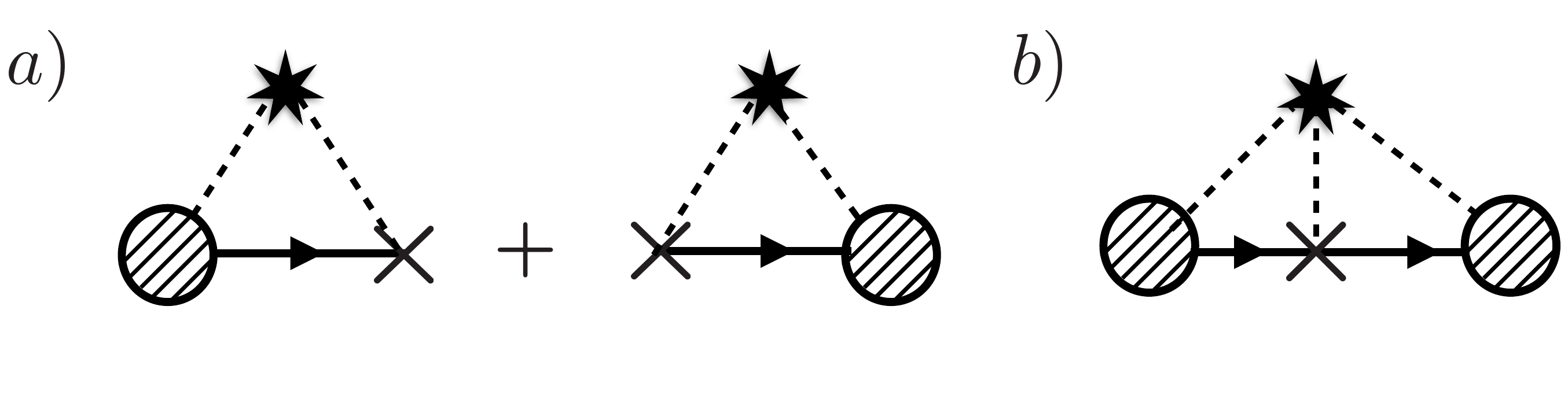}
\caption{ The spin-charge coupling self-energy at linear order in spin-orbit coupling strength $\lambda^2p_F^{2}$. There are two distinct class of Feynman diagrams: a) SOC vertex at the corner and b) SOC vertex in the middle. This has important consequences as discussed in the last paragraph of Sec.~\ref{sec:SCC}.
The shaded ``blob'' vertex stands the full Born series induced by spin-independent part of the impurity potential, while the cross represents the SOC vertex, see Fig.~\ref{fig:sigma0}.
 \label{fig:sigma1}
}
\end{figure}

The linear in SOC  self-energy,  $\check{\Sigma}^{(1)}$, is constructed from the vertex $t_{so}$ of Eq.~(\ref{t-so}) and the $8\times8$ zeroth order T-matrix defined in Eq.~\eqref{eq:fullT0}. This results in two different class of diagrams shown in Fig.~\ref{fig:sigma1}. 
The self-energy in Fig.~(\ref{fig:sigma1}a), which we defined as $\check{\Sigma}^{1a}$ can be analytically read off  as follow:
\begin{align}\label{eq:sigma000}
\check{\Sigma}^{1a}( \mathbf{r_1}, \mathbf{r_2}) 
=& \langle t_{so}(\vec{r}_1)\check{G}(\vec{r}_1,\vec{r}_2) \check{T}^{(0)}(\mathbf{r_2}) \rangle_{\mathrm{imp}} \nonumber \\
+& \langle \check{T}^{(0)}( \mathbf{r_1}) \check{G} (\vec{r}_1,\vec{r}_2) t_{so}^{\dagger}(\vec{r}_2) \rangle_{\mathrm{imp}}.
\end{align}
For short-range randomly distributed impurities, the correlation functions needed for performing the impurity average in these diagrams are
\begin{align} \label{eq:imp-ave}
&\langle t_{0}(\mathbf{r_{2}})
 \check{T}^{(0)}(\vec{r}_1) \rangle_{\mathrm{imp}}=\langle t_{0}(\mathbf{r_{1}})
 \check{T}^{(0)}(\vec{r}_2) \rangle_{\mathrm{imp}} 
 \nonumber \\
 &=n_{\mathrm{im}}  t_{0}(\mathbf{r_{2}})
 \check{T}^{(0)}(\vec{r}_1) \delta(\vec{r}_1-\vec{r}_2).
\end{align}
This closely resembles the correlators used in the calculations within the standard Born approximation 
$\langle t_{0}(\mathbf{r_{2}}) t_0(\vec{r}_1) \rangle_{\mathrm{imp}}=n_{\mathrm{im}} t_{0}^2 \delta(\vec{r}_1-\vec{r}_2)$. However, there are two important differences. Firstly, unlike the scalar $t_{0}$, the full scattering amplitude $\check{T}^{(0)}(\vec{r})$ is a $8\times 8$ matrix that depends on the GF $\check{g}$. Secondly, in our general setting we allow for a spatially varying disorder so that both $\check{T}^{(0)}(\vec{r})$ and $t_0(\vec{r})$ may depend on   spatial argument even after impurity average.
Next, we substitute  Eq.~(\ref{eq:fullT0}) and \eqref{eq:imp-ave}  into Eq.~\eqref{eq:sigma000} and found that it can be conveniently
decompose   as follow:
\begin{align} \label{eq:sigma_1a}
\check{\Sigma}^{1a} =   \mathbb{\check{\Sigma}}^{o}+   \mathbb{\check{\Sigma}}^{e}.
\end{align}
Here $\mathbb{\check{\Sigma}}^{o}$ ($\mathbb{\check{\Sigma}}^{e}$) is a series that contains
  odd (even) power of the  isotropic GF, $\check g $ introduced in Eq.~(\ref{eq:ansatz}). Note $\mathbb{\check{\Sigma}}^{o}$ ($\mathbb{\check{\Sigma}}^{e}$) corresponds to substituting the first (second) term  in   Eq.~(\ref{eq:fullT0}) into   Eq.~\eqref{eq:sigma000}.
To proceed further, we perform the Fourier transform with respect to the relative coordinate to go to the Wigner representation. As detailed in Appendix~\ref{app:sigma}, the result takes the following form:
\begin{widetext}
\begin{align}\label{eq:sigma-0}
\mathbb{\check{\Sigma}}^{o}(\vec{n},\mathbf{r})= & \frac{\omega_{1}\epsilon_{ajk}}{2}\bigg(-\frac{1}{3}n_{j}\left[\check{g}_{k} ,\sigma^{a}\right]-\frac{1}{2p_{F}}n_{j}\left\{ i\partial_{k}\check{g} ,\sigma^{a}\right\} 
+\frac{1}{6p_{F}}i\partial_{k}\left\{ \check{g}_{j} ,\sigma^{a}\right\} \bigg)  -\frac{i\epsilon_{ajk}\partial_{j}\omega_{1}}{12p_{F}}\{\sigma^{a},\check{g}_{k} \} 
\\
\mathbb{\check{\Sigma}}^{e}(\vec{n},\mathbf{r})= &
 \frac{-i\epsilon_{ajk}\omega_{2}}{2}\bigg(
\frac{1}{3}n_j \left(\sigma^a \check{g}_k \check{g} -\check{g}\check{g}_k \sigma^a\right)
-\frac{n_j}{2p_F}\left( \sigma^a (i\partial_k \check{g}) \, \check{g}+ \check{g}\, (i\partial_k \check{g} )\sigma^a
\right)
+\frac{i\partial_k}{6p_F} 
\left( 
\sigma^a  \check{g}_j\, \check{g} +\check{g} \, \check{g}_j \sigma^a
\right)
 \bigg)\nonumber \\
 &- \frac{\epsilon_{ajk}\partial_{j}\omega_{2}}{12p_F}\bigg(\sigma^{a}\,\check{g}_{k} \,\check{g} +\check{g} \,\check{g}_{k} \,\sigma^{a}\bigg) \label{eq:sigma_e}
\end{align}
\end{widetext}
Here $\omega_{1} $ and $\omega_{2}$ are spatially dependent scattering rates induced by SOC:
\begin{align} 
\omega_1 ( \mathbf{r})&= 2\pi n_{\mathrm{im}} N_F \mathrm{Re} \big[ t_{F}^*( \mathbf{r}) B( \mathbf{r})\big], \\
\omega_2 ( \mathbf{r})&= 2\pi n_{\mathrm{im}} N_F \mathrm{Im} \big[ t_{F}^*( \mathbf{r}) B( \mathbf{r})\big],
\end{align}
where $B(\vec{r})=\lambda^2 p_F^2 t_0( \mathbf{r})$ and $t_F$ is defined in Eq.~\eqref{eq:opsa}.  Notice that  $\omega_2$ is related to $\omega_1$ by the optical theorem $\omega_2(\vec{r})=\omega_1({\vec{r}})p_F a(\vec{r})$.
This form of parameterization is commonly used in discussing the  extrinsic spin-charge coupling in normal metallic state, see Ref.~\onlinecite{lifshitz2009} and Ref.~\onlinecite{chunli2016} for the discussion in 3D and 2D respectively. 
As mentioned earlier, the self-energy in Eq.~\eqref{eq:sigma-0} and \eqref{eq:sigma_e} have terms that are proportional to different powers of small parameters $\epsilon_1$ and $\epsilon_2$. For example, in Eq.~\eqref{eq:sigma-0}, the first, second, third and forth terms are of the order of $\epsilon_1$, $\epsilon_2$, $\epsilon_1\epsilon_2$ and $\epsilon_1\epsilon_2$ respectively.


Note that the last terms in Eq.~\eqref{eq:sigma-0} and Eq.~\eqref{eq:sigma_e} become crucial near material boundaries or interface of two materials where the gradient of the scattering rates, $\partial_j \omega_1$ and $\partial_j \omega_2$ are significant. In fact, the current induced spin-orbit torque that drives spin diffusion in metals is precisely induced at the material boundary \cite{abanin2009nonlocal,chunli2017}.
%
Within our quasiclassical approach, the spatial variation of the scattering rates  has to be over distances much larger than the Fermi wave length.

An important observation is that the spin-dependent corrections to the self-energy, Eqs.~\eqref{eq:sigma-0} and \eqref{eq:sigma_e}, contain terms that are  proportional to spatial derivative of the GF. These nonlocal terms give rise to $(p_F l)^{-1}$ correction to the spin-Hall angle and spin swapping coefficient. In the momentum space, these terms stem from the thin shell of the relative momentum near the Fermi surface. Hence, they are not captured in the equation of motion for the diagonal (in momentum space) component of the density matrix \cite{PhysRevB.81.125332,chunli2016}.

Let us now consider the self-energy coming from the diagram presented in Fig.~\ref{fig:sigma1}b. The corresponding analytic expression takes the form 
\begin{widetext}
\begin{align}
\check{\Sigma}^{1b}(\vec{r}_{1},\vec{r}_{2}) & =\int d\vec{r}_{3}\langle\check{T}^{(0)}(\vec{r}_{1})\,\check{G}(\vec{r}_{1},\vec{r}_{3})\,t_{so}(\vec{r}_{3})\,\check{G}(\vec{r}_{3},\vec{r}_{2})\check{T}^{(0)}(\vec{r}_{2})\rangle_{\text{imp}}\nonumber \\
  & =i\lambda^{2}\epsilon_{ajk}\int d\vec{r}_{3}\,\langle\check{T}^{(0)}(\vec{r}_{1})[\partial_{3j}\check{G}(\vec{r}_{1},\vec{r}_{3})]t_{0}(\vec{r}_{3})\sigma^{a}\partial_{3k}\check{G}(\vec{r}_{3},\vec{r}_{2})\,\check{T}^{(0)}(\vec{r}_{2})\rangle_{\text{imp}}\label{eq:sigma1b-def}
\end{align}
\end{widetext}
In the second line, we substituted the SOC vertex $t_{so}$ from Eq.~(\ref{t-so}) and performed the partial integration over $\vec{r}_3$ to move the derivative $\partial_{3j}$ to $\check{G}(\vec{r}_{1},\vec{r}_{3})$. To proceed further, we let $K= [\partial_{3j}\check{G}(\vec{r}_{1},\vec{r}_{3})] \sigma^{a}[\partial_{3k}\check{G}(\vec{r}_{3},\vec{r}_{2})]$ as  a matrix valued function and note that it is independent of the impurity position. Then, we perform the impurity average in Eq.~\eqref{eq:sigma1b-def} as follow:
\begin{align}
& \langle \check{T}^{(0)}(\vec{r}_{1})  t_0({\vec{r}}_3) \,K\,
\check{T}^{(0)}(\vec{r}_{2})\rangle _{\text{imp}} \nonumber \\
&= n_{\mathrm{im}} \delta(\vec{r}_{1}-\vec{r_{2}})\delta(\vec{r}_{3}-\vec{r}_{2}) \, \check{T}^{(0)}(\vec{r}_{1}) \,  t_0(\vec{r}_3)\,K\,
\check{T}^{(0)}(\vec{r}_{2}) \nonumber   \\
&=n_{\mathrm{im}} \delta(\vec{r}_{1}-\vec{r_{2}})\delta(\vec{r}_{3}-\vec{r}_{2}) \, \check{T}^{(0)}(\vec{r})  t_0(\vec{r})\,K\,
\check{T}^{(0)}(\vec{r} ) \label{eq:imp-avg3}
\end{align}
In the last line, using the Delta functions, we set $\check{T}^{(0)}(\vec{r}_{1}) = \check{T}^{(0)}(\vec{r}) $, $\check{T}^{(0)}(\vec{r}_{2}) = \check{T}^{(0)}(\vec{r}) $, and $t_{0}(\vec{r}_3)=t_0({\vec{r}_2})= t_0(\vec{r})$ where  $\vec{r}=(\vec{r_1}+\vec{r}_2)/2$.
Next, we substitute Eq.~\eqref{eq:imp-avg3} into Eq.~\eqref{eq:sigma1b-def} and perform the spatial integral (of $\vec{r}_3$) to arrive at a self-energy that is purely  local in space:
\begin{equation}
 \label{eq:sigma1b-1}
\check{\Sigma}^{1b}(\vec{r}_{1},\vec{r}_{2})=\delta(\vec{r}_1-\vec{r}_2)\check{\Sigma}^{1b}(\vec{r})\;,
\end{equation}
where  
\begin{equation}\label{eq:sigma1b-2}
\check{\Sigma}^{1b}(\vec{r})=in_{\rm im}\lambda^2t_{0}(\vec{r})\check{T}^{(0)}(\vec{r})\,\mathcal{K}(\vec{r})\,\check{T}^{(0)}(\vec{r})
\end{equation}
with the function $\mathcal{{K}}(\vec{r})$ defined as follow:
\begin{equation}
\mathcal{K}(\vec{r})=\epsilon_{ajk}\big[\partial_{3j}\check{G}(\vec{r},\vec{r}_{3})
\sigma^{a}\partial_{3k}\check{G}(\vec{r}_{3},\vec{r})\big]_{\vec{r}_3=\vec{r}}.
\end{equation}
Note that  $\check{\Sigma}^{1b}(\vec{r})$ in Eq.~(\ref{eq:sigma1b-1}) is, in fact,  the self-energy in the Wigner representation required for the quasiclassical collision integral.  
Now we substitute the T-matrix from Eq.~(\ref{eq:fullT0}) and use the following identities to evaluate the function $\mathcal{{K}}(\vec{r})$ entering Eq.~(\ref{eq:sigma1b-2}):
\begin{align}
 \big[\partial_{3j}\check{G}(\vec{r},\vec{r}_{3})\big]_{\vec{r}_3=\vec{r}}=& 
-i\pi N_F\left[\frac{1}{2}\partial_j\check{g}(\vec{r})-i\frac{p_F}{3}\check{g}_j(\vec{r})\right],
\\
\big[\partial_{3k}\check{G}(\vec{r}_{3},\vec{r})\big]_{\vec{r}_3=\vec{r}}=& 
-i\pi N_F\left[\frac{1}{2}\partial_k\check{g}(\vec{r})+i\frac{p_F}{3}\check{g}_k(\vec{r})\right].
\end{align}
These identities are derived by using the Wigner representation of $\check{G}(\vec{r},\vec{r}_{3})$ together with the definitions of the quasiclassical GF and its first two moments, $\check{g}$ and $\check{g}_k$. Finally, after straightforward algebra, Eq.~(\ref{eq:sigma1b-2}) reduces to the following form

\begin{align}
\check{\Sigma}^{1b}\left(\mathbf{r}\right)=&i\, n_{\rm im}\, \lambda^{2}t_{0}(\pi N_{F}p_{F})^{2}\Big[-(\text{Re\,}t_{F})^{2}\tilde{\mathcal{K}} +\nonumber \\
&(\text{Im\,}t_{F})^{2}\check{g}\tilde{\mathcal{K}}\check{g} -(\text{Im\,}t_{F}\text{Re\,}t_{F})\big(\check{g}\tilde{\mathcal{K}}+\tilde{\mathcal{K}}\check{g}\big)\Big],\label{eq:sigma1b-fin}
\end{align}
where
\begin{equation}
\tilde{\mathcal{K}}=\epsilon_{ajk}\Big[\frac{1}{9}\check{g}_{j}\sigma^{a}\check{g}_{k}+\frac{i}{6p_{F}}\partial_{j}\check{g}\sigma^{a}\check{g}_{k}-\frac{i}{6p_{F}}\check{g}_{j}\sigma^{a}\partial_{k}\check{g}\Big].\label{tildeK}
\end{equation}
In the above expressions we neglect  the term proportional to the product $\partial_{j}\check{g}\partial_{k}\check{g}$ as it is of the order of $\epsilon_2^2$. The latter is beyond our accuracy corresponding to retaining only terms linear in $\epsilon_2$ and at most proportional to $\epsilon_1^2$ or $\epsilon_1\epsilon_2$, which exactly corresponds to the terms kept in Eq.~(\ref{tildeK}).

In contrast to $\check{\Sigma}^{1a}$ in Eqs.~(\ref{eq:sigma_1a})-(\ref{eq:sigma_e}),  the self-energy $\check{\Sigma}^{1b}$ in Eqs.~(\ref{eq:sigma1b-fin}) does not depend on the external momentum. This has two important consequences for the derivation of the Usadel equation. Firstly, the anticommutator term in the zeroth moment $\check{\cal{I}}_0$ of the collision integral in Eq.~(\ref{eq:coll-inta}) vanishes for $\check{\Sigma}^{1b}$. Secondly, within our accuracy $\check{\Sigma}^{1b}$ does not contribute to the first moment $\check{\cal{I}}_k$ of the collision integral defined by Eq.~(\ref{eq:coll-intb}). As $\check{\Sigma}^{1b}$ by itself contains terms proportional to $\epsilon_1^2$ and $\epsilon_1\epsilon_2$ it brings to $\check{\cal{I}}_k$ the corrections of the order of $\epsilon_1^3$ and $\epsilon_1^2\epsilon_2$ which are irrelevant in our diffusive limit. 
Exactly the same arguments apply to the part of $\check{\Sigma}^{1a}$ that does not depend on the external momentum $\vec{n}$. Therefore the self-energy $\check{\Sigma}^{1b}$ and the $\vec{n}$-independent part of $\check{\Sigma}^{1a}$ contribute only to the first term in $\check{\cal{I}}_0$ of Eq.~(\ref{eq:coll-inta}), while the $\vec{n}$-dependent part of $\check{\Sigma}^{1a}$ gives nonvanishing contributions to both $\check{\cal{I}}_0$ and  $\check{\cal{I}}_k$. We will make use of these properties later  in Secs.~\ref{sec:g_k} and \ref{sec:Usadel}.
 
 Before we close this subsection, let us emphasize the need to retain the Poisson bracket in Eq.~\eqref{eq:coll-inta}. As shown in Eqs.~\eqref{eq:sigma-0},~\eqref{eq:sigma_e} and \eqref{eq:sigma1b-fin}, $\check{\Sigma}^{1a} $ and  $\check{\Sigma}^{1b} $ are made up by terms of the order $\epsilon_1$, $\epsilon_2$ and  $\epsilon_1\epsilon_2$. 
 When we substitute all of them 
  into Eq.~\eqref{eq:coll-inta} to evaluate the collision integral,  terms of order  $\epsilon_1$  in the self-energy
 can enter the Poisson bracket and generate terms of  order  $\epsilon_1 \epsilon_2$ since the Poisson bracket involves spatial derivative. This is precisely the result one would get from substituting order $\epsilon_1 \epsilon_2$ terms of the self-energy into the commutator in Eq.~\eqref{eq:coll-inta}. Hence, in order to derive the Usadel equation in the presence of SOC correctly, it is necessary to retain the Poisson bracket in Eq.~\eqref{eq:coll-inta}.

\subsubsection{$\check{\Sigma}^{(2)}$: Spin-relaxation}
The last self-energy we consider is of the second order in the SOC strength. We only include the leading order  term in the counting scheme which describes spin relaxation:
\begin{equation} \label{eq:sigma_EY}
\check{\Sigma}^{(2)}(\vec{n}, \mathbf{r})=\frac{16 i}{3 \tau_{\mathrm{so}} (\vec{r})}  \left(\sigma^i n_i \check{g}  \sigma^j n_j - \sigma^a \check{g}  \sigma^a \right),
\end{equation}
Here the Elliott-Yafet spin relaxation time is given by
\begin{equation}
\frac{1}{\tau_{\mathrm{so}}(\vec{r}) }= \frac{8}{9\tau(\vec{r})} \lambda^4 p_F^{4}
\end{equation}

It is worth mentioning that most of the works concerning the effect of extrinsic SOC in  superconducting state \cite{soi_AG,demler1997superconducting,bergeret2007scattering} focus only on the spin relaxation described by our $\check{\Sigma}^{(2)}$ while the self-energy related to the spin-charge coupling $\check{\Sigma}^{(1)}$ was disregarded.

 \subsection{Constitutive relation and the normalization condition} \label{sec:g_k}
After computing all the self-energies, the next step in the derivation of the Usadel equation is to eliminate the first moment $\check{g}_k$ in Eq.~(\ref{eq:g_k}) by expressing it in terms of the zeroth moment $\check{g}$. In the following, we call the map $\check{g}\mapsto\check{g}_k$ the constitutive relation as it is similar to the relation between the current and the density in the usual diffusion theory. By substituting Eq.~(\ref{eq:ansatz}) into Eq.~(\ref{eq:coll-intb}) we rewrite the collision integral $\check{\cal{I}}_k$ entering Eq.~(\ref{eq:g_k}) more explicitly as follows 
\begin{equation}\label{I_k-gen}
 \check{\cal{I}}_k = -\frac{i}{3}[\langle\check{\Sigma}\rangle,\check{g}_k] -i[\langle n_k\check{\Sigma}\rangle,\check{g}].
\end{equation}

As we argued in the discussion after Eqs.~(\ref{eq:sigma1b-fin}) and (\ref{tildeK}), the self-energy $\check{\Sigma}^{(1)}$ gives  negligible (of the order of $\epsilon_1^3$ and $\epsilon_1^2\epsilon_2$) contributions to the first term in Eq.~\eqref{I_k-gen}. Therefore only the Drude self-energy $\check{\Sigma}^{(0)}$ contributes to $\langle\check{\Sigma}\rangle$ in Eq.~(\ref{I_k-gen}).
For the same reason,  $\langle n_k\check{\Sigma}\rangle$ in the second term of Eq.~(\ref{I_k-gen}) is fully determined by the $\vec{n}$-dependent part of $\check{\Sigma}^{1a}$; those $\vec{n}$ independent part of $\check{\Sigma}^{1a}$ contributes to the collision integral at order $\epsilon_1^3$ and $\epsilon_1^2\epsilon_2$. Thus,  Eq.~(\ref{eq:g_k}) leads to  the constitutive relation:
\begin{equation}
 \label{eq:g_k-fin}
\frac{v_{F}}{3} \partial_k \check{g} = -\frac{i}{3}[\check{\Sigma}^{(0)},\check{g}_k] -i[\langle n_k\check{\Sigma}^{1a}\rangle,\check{g}].
\end{equation}
By substituting Eqs.~\eqref{eq:sigma_0} and \eqref{eq:sigma_1a} into this equation, and rearranging the terms we bring it to the following compact form,
\begin{equation} \label{eq:norm-presv}
 \frac{v_{F}}{3} \partial_k \check{g} + \big[ \check{\mathcal{A}}_k \, , \,  \check{g} \big] =0,
\end{equation}
where $\check{\mathcal{A}}_k$ is given by the following expression,
\begin{align}  
\check{\mathcal{A}}_k &= \frac{\check{g}_k}{6\tau} + \frac{\omega_1 \epsilon_{ajk}}{6} \left( \frac{1}{3} \,i \left[\check{g}_{j} ,\sigma^{a}\right] - \frac{1}{2p_{F}} \left\{ \partial_{j}\check{g} ,\sigma^{a}\right\} \right) \nonumber \\
&+ \frac{\omega_2 \epsilon_{ajk}}{12p_F}   \left(   \sigma^a i\partial_j \check{g}\, \check{g} + \check{g}\,i \partial_j \check{g} \sigma^a \right) \nonumber \\
&+ \frac{\omega_2 \epsilon_{ajk}}{18}   \left( \sigma^a \check{g}_j \check{g}    - \check{g}\, \check{g}_j \sigma_a \right).\label{eq:Akk}
\end{align}
The first term in the above expression comes from the first (Drude) term in the right hand side of Eq.~(\ref{eq:g_k-fin}) while the rest corresponds to the second term in Eq.~(\ref{eq:g_k-fin}) and arises from the parts proportional to the external momentum $n_k$ in Eqs.~\eqref{eq:sigma-0} and  \eqref{eq:sigma_e}. 

Importantly, the structure of Eq.~\eqref{eq:norm-presv} suggests that $\partial_k \check{g}^2=0$ and this allows us to impose the standard normalization condition on the Usadel Green function:
\begin{equation}
\check{g}^2 = 1.
\end{equation} 

To proceed further, we expand $\check{g}_k$ to linear order in SOC: $\check{g}_k=\check{g}_k^{(0)}+ \check{g}_k^{(1)}$ where $\check{g}_k^{(1)}\propto (\lambda p_F)^{2}$. At  zeroth order in SOC, Eq.~\eqref{eq:norm-presv} reads,
\begin{equation}
\frac{v_{F}}{3}\partial_{k}\check{g} =    -\frac{1}{6\tau }\left[\check{g} \,,\,\check{g}_{k}^{(0)} \right].
\end{equation}
Because of the normalization condition $\check{g}^2 = 1$, this equation leads to the well-known solution  
\begin{equation}
 \label{g_k-0}
\check{g}_{k}^{(0)}=-l \check{g} \partial_k \check{g}
\end{equation}
where $l=v_F \tau$ is the mean free path. At the linear in SOC order we substitute the zeroth order solution $\check{g}_{k}^{(0)}$ into the terms proportional to $\omega_1$ and $\omega_2$ in Eqs.~(\ref{eq:norm-presv}-\ref{eq:Akk}). This generates the following equation for the linear in $\lambda^2 p_F^2$ correction $\check{g}_k^{(1)}$,
\begin{equation} \label{eq:temp1}
\bigg[  \check{g}\, , \,
\frac{\check{g}_{k}^{(1)}}{6\tau} - \frac{v_F \epsilon_{akj}}{12}  
\left(  \theta'  \left\{\partial_{j}\check{g} ,\sigma^{a}\right\} 
+i \kappa'  \left[\check{g} \partial_{j}\check{g} ,\sigma^{a}\right] \right)
 \bigg]=0,
\end{equation}
where the parameters $\theta'$ and $\kappa'$ are defined as follows
\begin{align}
\theta' &=  \frac{2}{3}\omega_2 \tau  + \frac{\omega_1  \tau }{p_F l }, \label{eq:l_H} \\
\kappa' &=\frac{2}{3}\omega_1 \tau  -\frac{\omega_2  \tau }{p_Fl }  . \label{eq:l_kappa}
\end{align}
The solution that satisfies Eq.~(\ref{eq:temp1}) is given by, 
\begin{equation}\label{g_k-1}
\check{g}_{k}^{(1)}=\frac{l }{2} \epsilon_{akj} \big(  
 \theta'  \left\{\partial_{j}\check{g} ,\sigma^{a}\right\} 
+i \kappa'  \left[\check{g} \partial_{j}\check{g} ,\sigma^{a}\right]
 \big)
\end{equation}
%
 We combine the results of   Eqs.~(\ref{g_k-0}) and (\ref{g_k-1}) and find the final expression that relates $\check{g}_k$ to $\check{g}$,
\begin{align} \label{eq:bare-current}
&\frac{v_F}{3}\check{g}_{k}  =  -D  \check{g}  \partial_{k}\check{g}    
+ \frac{D  }{2} \epsilon_{akj} \bigg(  \theta'  \left\{\partial_{j}\check{g} ,\sigma^{a}\right\} 
+i \kappa'  \left[\check{g} \partial_{j}\check{g} ,\sigma^{a}\right] \bigg).
\end{align}
This equation is the constitutive relation that we need for the derivation of a  closed  equation for the isotropic GF $\check{g}$.
Typically the relations of this sort establish a connection between the diffusion ``current'' and the density. However, due to the presence of SOC, the first moment of the GF, $\check{g}_k$, is \emph{not} the conserved current entering the continuity equation . This is because SOC depends on the particle's momentum and it produces an additional contribution to the current, the  so called anomalous current. At the level of the kinetic theory, the anomalous current comes from \textit{both} the commutator and anticommutator in the collision integral,  Eq.~\eqref{eq:coll-inta}. 

 It is worth mentioning that $\theta' $ and $\kappa'$ defined by Eqs.~\eqref{eq:l_H} and (\ref{eq:l_kappa}) are \textit{not} the total spin Hall angle and the spin swapping coefficient. In Eq.~\eqref{eq:l_H}, the first term  is the skew-scattering contribution while the second term is only \textit{half} of the side-jump contribution to the spin Hall angle. The other half comes from the anomalous current, exactly as it happens at the level of the Born approximation \cite{Raimondi2012}. Similarly, for the spin swapping coefficient $\kappa$ the second term in Eq.~\eqref{eq:l_kappa} will be doubled due to the anomalous contribution. We will return to this discussion in the next subsection after completing the derivation of the Usadel equation.

\subsection{Usadel equation} \label{sec:Usadel}

Let us substitute the constitutive relation, Eq.~(\ref{eq:bare-current}) into Eq.~\eqref{eq:g_0} and obtain the generalized Usadel equation. In the absence of SOC, we recover the usual Usadel equation by 
setting the right hand side in Eq.~\eqref{eq:g_0} to zero,  and by using the zeroth order constitutive relation,  Eq.~(\ref{g_k-0}) on the left  hand side. 
To second order in SOC, the Usadel equation reads:
\begin{align} \label{eq:USA}
&\tau_{3}\partial_{t_1}\check{g} + \partial_{t_2}\check{g} \tau_{3} +\frac{v_F}{3} \partial_{k}\check{g}_k 
+i\left[\check{\Delta}  ,\check{g} \right]
=\check{\mathcal{I}}_0^{(1)}+\check{\mathcal{I}}_0^{(2)},
\end{align}
where $\mathcal{\check{I}}^{(2)}_0 $ describes the standard (Elliott-Yafet) spin relaxation \cite{soi_AG,demler1997superconducting,bergeret2007scattering} and is obtained by  substituting $\check{\Sigma}^{(2)}$,  Eq.~\eqref{eq:sigma_EY},  into the first term of Eq.~\eqref{eq:coll-inta}: 
\begin{equation} \label{eq:EY}
\mathcal{\check{I}}^{(2)}_0 =- \frac{1}{8\,\tau_{\mathrm{so}} } \left[\sigma^a \check{g}  \sigma^a ,  \check{g}  \right].
\end{equation}

 The evaluation of $\check{\mathcal{I}}_0^{(1)}$ is more cumbersome  and requires some care. One  has to substitute the first order self-energies  $\check{\Sigma}^{1a}$,  Eqs.~\eqref{eq:sigma_1a}-(\ref{eq:sigma_e}), and $\check{\Sigma}^{1b}$,  Eqs.~(\ref{eq:sigma1b-fin})-(\ref{tildeK}), into Eq.~\eqref{eq:coll-inta}. Note that both the commutator and the anticommutator 
\footnote{The momentum derivative in  the self-energy is understood as $\partial_{p_i} \check{\Sigma}^{(1)}(\mathbf{p},\mathbf{r})=(p_F)^{-1}\partial_{n_i} \check{\Sigma}^{(1)}(\mathbf{n},\mathbf{r})$.}
terms in Eq.~\eqref{eq:coll-inta} contribute to $\mathcal{\check{I}}^{(1)}_0$. After some lengthy algebra detailed in Appendix \ref{app:coll-int}, the collision integral $\mathcal{\check{I}}^{(1)}_0$ can be represented compactly as sum of two distinct contributions,
%
%
%
\begin{equation}  \label{eq:coll-int-g0}
\mathcal{\check{I}}^{(1)}_0 = \check{\cal{T}}  -\partial_k \check{\mathcal{J}}^{\mathrm{an}}_k
\end{equation}
The the $8\times8$ matrix $\check{\cal{T}}$ and the matrix-valued vector $\check{\mathcal{J}}^{\mathrm{an}}_k $ are defined as follows,
\begin{widetext}
\begin{align}
\check{\mathcal{J}}^{\mathrm{an}}_k =  \frac{D}{2} \epsilon_{akj}    \bigg[ \frac{\omega_1  \tau}{p_{F} l }\,\big\{\partial_{j}\check{g} ,\sigma^{a}\big\}+ i \frac{\omega_2 \tau}{p_Fl }  [\sigma^a, \check{g} \partial_j \check{g} ]\bigg],
\end{align}
\begin{align}
\check{\cal{T}}=\frac{D}{4}\epsilon_{akj} &\bigg[
 \left( \frac{2}{3} \omega_2 \tau + \frac{2 \omega_1 \tau  }{p_F l } \right) 
\,\big[\sigma^{a},\check{g} \partial_{k}\check{g} \partial_{j}\check{g} \big] 
+\left(\frac{2}{3}\omega_1\tau -\frac{2\omega_2 \tau}{p_Fl}  \right)
i\big[\partial_{k}\check{g}\partial_{j}\check{g},\sigma^{a}] \bigg].
\end{align}
\end{widetext}
Due to the way the above quantities enter the diffusion equation, we identify $\check{\mathcal{J}}^{\mathrm{an}}_k $ as the  \emph{anomalous current}  and  $\check{\cal{T}}$ as the \emph{spin-orbit-torque}.

It is important to emphasize that all the kinetic coefficients, $D,\omega_1 ,\omega_2, \tau, l$ depend on the spatial coordinate $\vec{r}$.
 Therefore, one cannot redefine the spin-torque and anomalous current by simply absorbing part of $\check{\cal{T}}$  into $\partial_k \check{\mathcal{J}}^{\mathrm{an}}_k $ or vice versa. In other words, the definition of  $\check{\cal{T}}$ and $\partial_k \check{\mathcal{J}}^{\mathrm{an}}_k $ is unambiguous when we allow the kinetic coefficients to vary in space.
\footnote{There is still a ``trivial'' curl ambiguity of the current itself $\check{\mathcal{J}}_k^{\mathrm{an}} \rightarrow \check{\mathcal{J}}_k^{\mathrm{an}} + \epsilon_{kjl}\partial_j \check{O}_l$ where $\check{O}_l$ can be any vector. This ambiguity is not of our concern here as we are mostly interested in global flows conserved across extended surfaces.} 
Next, we move the total divergence of the anomalous current (i.e.~$-\partial_k \mathcal{J}_k^{\mathrm{an}}$) to the left hand side of Eq.~(\ref{eq:USA})  and define the total current as the sum of the first moment $v_F\check{g}_{k}/3$ ({\it cf.}~Eq.~\ref{eq:bare-current}) and  $\mathcal{J}_{k}^{\mathrm{an}}$:
\begin{align} \label{final_current}
\check{\cal{J}}_k=& \frac{v_F}{3} \check{g}_k + \check{\mathcal{J}}^{\mathrm{an}}_k    \nonumber \\
=&   -D  \check{g}  \partial_{k}\check{g}    
+ \frac{D  }{2} \epsilon_{akj} \big(  \theta  \left\{\partial_{j}\check{g} ,\sigma^{a}\right\} 
+i \kappa  \left[\check{g} \partial_{j}\check{g} ,\sigma^{a}\right] \big).
\end{align}
Here the \textit{total} spin Hall angle and spin swapping coefficient are given by the following expressions, 
\begin{equation} \label{eq:theta}
\theta =\theta' +\frac{\omega_1 \tau }{p_F l }\equiv  \frac{2}{3}\omega_2 \tau  + 2\frac{\omega_1  \tau }{p_F l },
\end{equation}
\begin{equation} \label{eq:kappa}
\kappa =\kappa' +\frac{\omega_2  \tau }{p_Fl }\equiv \frac{2}{3}\omega_1 \tau  -2\frac{\omega_2  \tau }{p_Fl } .
\end{equation}
Equation \eqref{final_current} is the result announced in Eq.~(\ref{eq:Jk}) of Sec.~\ref{sec:result}. Interestingly, the term describing the spin Hall effect in Eq.~\eqref{final_current}, {\it i.e.~}the term proportional to $\theta$,  has exactly the same form as the spin Hall term obtained in superconductors with intrinsic spin-orbit coupling\cite{tokatly2017usadel}. This means that in systems with both extrinsic and intrinsic SOC the coefficient describing the charge-spin coupling is simply the sum of the two contributions.

As we have already discussed in the previous subsection, the first (second) term in Eq.~(\ref{eq:theta})  corresponds to the skew-scattering ( side-jump) contribution to  the spin Hall angle $\theta$. One half of the side-jump contribution comes from the SOC correction to the anisotropic part of the GF $\check{g}_k^{(1)}$ while another half appears from the anomalous current.

Similarly, the spin swapping coefficient $\kappa$ also receives contributions from two independent scattering mechanisms. 
The first term in Eq.~\eqref{eq:kappa} exactly reproduces the swapping coefficient identified by Lifshitz and Dyakonov \cite{lifshitz2009}. In addition we found another contribution given by the second term in Eq.~\eqref{eq:kappa} which is proportional to $(p_Fl)^{-1}$ and relies on the nonlocality of the self-energy. This term  gives rise to a ``non-local'' component to the spin current swapping effect, which is formally similar to the side-jump component of the spin Hall 
effect. In fact the one half of the ``non-local'' contribution to  $\kappa$ comes from the ``normal'' and another half from the anomalous currents. 
Note also that the side jump contribution to $\theta$ and the ``non-local'' contribution to $\kappa$ scale in exactly the same way with respect to the impurity concentration -- both are proportional to $n_{\mathrm{im}}$ and thus inversely proportional to the Drude conductivity.

In a homogeneous system  where gradient of kinetic coefficients vanished $\partial_k \theta(\vec{r})=\partial_k \kappa(\vec{r}) =0$,  the divergence of the current (or the source of diffusion) takes the usual form $ \partial_k \check{\cal{J}}_k =-\partial_k \left(D\check{g} \partial_k \check{g}\right)$.   Hence, a flow (current) induced by SOC typically occurs at material boundaries and/or interface of two materials where $\partial_k \theta(\vec{r})$ and $\partial_k \kappa(\vec{r})$ are finite. 
Although our quasiclassical theory cannot describe  boundary effects  which occurs at  the scales smaller than the mean free path, it provides  an unambiguous definition of the generalized matrix current whose conservation define a boundary condition for the kinetic equations.
The same philosophy has been used extensively to model spin-charge conversion also in the normal state \cite{abanin2009nonlocal,PhysRevB.94.121408,chunli2017}. 
In the next section we compile all the results above and obtain the generalized Usadel equation.

\section{Discussion} \label{sec:discussion}
Substitution of  Eq.~\eqref{eq:bare-current}, \eqref{eq:EY} and \eqref{eq:coll-int-g0} into Eq.~\eqref{eq:USA} leads to the main result of our paper, the generalize  Usadel equation:
 \begin{equation} \label{eq:Usadel2}
 \tau_{3}\partial_{t_1}\check{g}+ \partial_{t_2}\check{g} \tau_{3} +i [\hat{\Delta} ,\check{g} ]+\partial_{k}\check{\cal{J}}_k
 =  \frac{-1}{8\tau_{\mathrm{so}}} \left[\sigma^a \check{g} \sigma^a ,  \check{g} \right]+ \check{\cal{T}}.
 \end{equation}
Recall $\check{g}=\check{g}(\vec{r},t_1,t_2)$ and
 the generalized current $\check{\cal{J}}_k$ is defined in Eq.~\eqref{final_current} and the SOC induced torque on the right hand side is given by  
\begin{align} \label{eq:torque2}
\check{\cal{T}}&=\frac{D}{4}\theta\,\epsilon_{akj}\,\big[\sigma^{a},\check{g}\partial_{k}\check{g}\partial_{j}\check{g}\big]+\frac{D}{4}\kappa\,\epsilon_{akj}\,i\big[\partial_{k}\check{g}\partial_{j}\check{g},\sigma^{a}] \; .
\end{align}

In order to discern the physics behind these expressions, it is helpful to first study  Eq.~(\ref{eq:Usadel2}) in a (normal state) metal. The advantage of using quasiclassical equation is that it describes both superconducting and normal state in a coherent manner. Indeed, the normal state diffusion equation can be readily obtained by setting the retarded and advanced GFs to $+\tau_3\delta(t_1-t_2)$, $-\tau_3\delta(t_1-t_2)$  respectively, and the time $t_1=t_2=t$ in the Keldysh  GF. Then,  from Eq.~(\ref{eq:Usadel2}), we can obtain the  well-known spin diffusion equation after multiplying it by the vector $\sigma$ and  taking the trace \footnote{For the present discussion about spin-charge diffusion in metal, it is sufficient to consider the $(1,1)$ element of the Green function in Nambu space.}
\begin{equation}
\label{eq:sde}
\partial_t S^a+\partial_{k}{{\cal J}}_k^a=\frac{1}{\tau_{so}}S^a\; , 
\end{equation}
where $S^a=-(\pi/4)N_F{\rm Tr}\,\sigma^a\tau_3\, \check g^K$ is the $a$-component of the \textit{non-equilibrium} spin density. It measures the deviation from the equilibrium spin-density.  The equilibrium spin density can induce, for example, from  a static  Zeeman field.
Similarly, one can obtain the diffusion equation for the charge density,  $n=-(\pi/4)N_F{\rm Tr}\, \check g^K$ $n$ by taking the trace over spin $\sigma$.

Notice that the  term  $\check{\cal{T}}$ on the  right hand side of Eq.~(\ref{eq:Usadel2}) does not contribute to the normal state spin diffusion equation, Eq.~\eqref{eq:sde}. In other words, the right hand side of Eq.~(\ref{eq:sde}) which describes spin torque contains only the  well-known Elliott-Yafet spin relaxation term.  Mathematically, the reason for the  vanishing $ \check{\cal{T}}$ contribution is due to the fact that this term contains  products of  two derivatives of GFs. In order to obtain the Keldysh component of $\check{\cal{T}}$, one necessarily  needs to  differentiate  at least one retarded or advanced GF which are constant  in space.   We will see that this  is different in the superconducting state where  $ \check{\cal{T}}\neq 0$.

Next, we discuss the equation of motion for the currents. By taking the trace over spin in Eq.~(\ref{final_current}), one  obtains   the charge current,   which in the normal state reads:
\begin{equation}
\label{eq:chJ_normal}
J_k=-D\partial_k n+D\epsilon_{akj}\theta\partial_j S^a \; .
\end{equation} 
The spin current is obtained  by multiplying  Eq.~(\ref{final_current}) with $\sigma^b$ and taking the trace:  
\begin{equation}
\label{eq:spinJ_normal}
J_k^b=-D\partial_k S^b+D\epsilon_{bkj}\theta\partial_j n-D\kappa\left(\delta_{kb}\delta_{jaa}-\delta_{ka}\delta_{jb}\right)\partial_j S^a  
\end{equation} 
The first term  on the  right hand side  of these  two equations are   the charge and spin   diffusion currents respectively.  The  terms  proportional to $\theta$ couple  charge and spin degrees of freedom and they describe the spin Hall effect and inverse spin Hall effect. They  stem from the anti-commutator in Eq.~(\ref{final_current}). The third term on the  right hand side of Eq. (\ref{eq:spinJ_normal}),   leads to the swap of the spin and direction indexes of the spin-current \cite{dyakonov_swapcurrent2009}.    
Thus,  the generalized  Usadel equation leads,  in the normal state, to the known effects regarding spin-charge coupling.   

Having established the known results in the normal phase, let us now discuss Eq.~(\ref{eq:Usadel2}) in  the superconducting phase  which  is characterized by non-trivial excitation spectrum, i.e.~the retarded and  advanced Green functions are  non-trivial functions of the time difference $t_1-t_2$  and of the space coordinate.  More importantly, their matrix structure in the  Nambu space (c.f.~Eq.~\ref{eq:qc_GF} ) carries  finite anomalous (off-diagonal) components  which leads to, among many other phenomena, the \textit{equilibrium} supercurrent and the \textit{equilibrium} magnetoelectric effect \cite{bergeret2016} which we will discuss below.
 In the normal metallic phase, the charge and spin current is related to different components of the gradients of charge and spin densities as in  Eqs.~(\ref{eq:chJ_normal}-\ref{eq:spinJ_normal}). However, this is not generally the case in superconductor due to the 
 the  non-trivial dependence of retarded and advance GFs on  energy.   
For example, while the second  (spin Hall)  term on the  right hand side of Eq. (\ref{final_current}) can indeed be written as a spin or charge (spectral) density, the third (swapping) term proportional to $\kappa$ cannot, due to the product $\check g\partial _k \check g$.

 We start by inspecting the expression for the $8\times8$ matrix current described in Eq.~(\ref{final_current}).
 The charge current is obtained by  taking trace in Eq.~(\ref{final_current}) after multiplying it with the Nambu matrix $\tau_3$.
 In the absence of SOC, only the first term in Eq.~(\ref{final_current}) is non-vanishing. Notice that  even in equilibrium, this term can result in a finite current  if the   anomalous GF $\check f$ and its time-reversal conjugate $\check f^c$ are different.
For instance, this can happen in a bulk superconductor with a finite phase gradient which leads to the well-known supercurrent. 
 Similarly, there exists equilibrium magnetoelectric effects  in superconductor with SOC, i.e. spin to charge conversion in the absence of spin-injection (pumping) field.  For example, a static Zeeman field can polarize  the condensate and  creates  triplet  correlations. These correlations enter the  second (SH) term in Eq.~(\ref{final_current}) and contribute to the charge current. Reciprocally, a charge supercurrent may induce a spin current in a superconductor.  These effects have been studied in Ref. \onlinecite{bergeret2016} within the first  Born approximation. 
 
Interestingly, the expression for the current, Eq.~(\ref{final_current}), has exactly the same form as in the Born approximation studied in  Ref.~\onlinecite{bergeret2016}. However, unlike  Ref.~\onlinecite{bergeret2016}, the analysis in the present  work is valid to all orders in the scalar elastic impurity potential and includes resonant skew scattering and side-jump mechanism. This results in the renormalization of the kinetic coefficients (spin Hall angle and swapping coefficient)   discussed in  Eqs.~(\ref{eq:theta})-(\ref{eq:kappa}).  

More importantly, we found a hitherto unknown term in the Usadel equation:  ${\cal \check{T}}$ in the  right hand side of Eq.~(\ref{eq:Usadel2}) and defined in Eq.~(\ref{eq:torque2}). Since ${\cal \check{T}}$ is a commutator with the spin Pauli matrix,  it vanishes under trace in spin-space. In other words, it does not modify the (spectral) charge diffusion equation and only enters the spin diffusion equation as a spin-orbit torque whose magnitude is characterized by the spin Hall angle and swap current coefficient.  As mentioned above, this torque vanishes in the normal state [{\it cf.} Eq. (\ref{eq:spinJ_normal})] hence it describes  a spin torque unique for the superconducting state. 

We now  illustrate  with an example the consequences of this new term   by considering  a superconductor with a single magnetic vortex.   The latter  is described by a spatial dependent  order parameter
$\Delta=|\Delta(r)|e^{in\varphi}$, where $r$ is the radial component of the position vector $\mathbf{r}$, $\varphi=\tan^{-1}(y/x)$ is the polar angle  and $n$ is the vorticity or the topological charge.   The axis of the vortex is in $z$-direction. 
Let us assume that the superconductor is subject to a homogeneous  spin-splitting (Zeeman) field  $-h^a\sigma^a$  where $h^a$ is the unit vector pointing along the direction of the Zeeman field. In the absence of SOC, the normal and anomalous components of the GF, Eq.~\eqref{eq:qc_GF}, have the following form:
\begin{align}
\hat g(r)&=g_s(r)+g_t(r)h^a\sigma^a \label{eq:ghansatz}\\
\hat f(r,\varphi)&=\left[f_s(r)+f_t(r)h^a\sigma^a\right]e^{in\varphi}\label{eq:fhansatz}\; .
\end{align}
 Note the $\varphi$ dependence enters only in the anomalous GF $ \hat f(r,\varphi)$. Next, we substitute Eq.~\eqref{eq:ghansatz} and \eqref{eq:fhansatz} into Eq.~(\ref{eq:Usadel2}) to 
compute the spin torque to linear order in SOC. The result reads:
\begin{equation}
{\rm Tr } \left[ \sigma \tau_3 \check {\cal T}\right]=F(r)\left(n\,\mathbf{z}\times \mathbf{h}\right).\label{eq:vortexT}
\end{equation}
Here the function $F(r)$ is a function of the radial components and can be computed by solving the Usadel equation, but its explicit form is not relevant for this discussion.  Due to  the Nambu matrix $\tau_3$ in the trace of Eq.~\eqref{eq:vortexT},  $F(r)$  contains products of the anomalous GF and its time-conjugated with derivatives with respect to $r$ and $\varphi$.   Therefore, this spin torque only appears in the superconducting state. For the example consider here the amplitude of the torque decays away from the vortex core, {i.e.} with  increasing $r$. 
From Eq.~\eqref{eq:vortexT}, we conclude that the spin-torque generated by $\check{\mathcal{T}}$ is proportional to the vector product between the angular momentum of the condensate $n\,\vec{z}$ (where  $n$ is the vorticity of the vortex) and the triplet vector $\vec{h}$.

Let us emphasis all the spin-charge conversion (or spin-orbitronic) effects we discussed in this article occurs at leading order in spatial non-uniformity ($\partial_i^2$) of the Usadel equation. The characteristic scale of the spin torque $\mathcal{\check{T}}$ is determined by the spin Hall angle $\theta$ and  swap current coefficient $\kappa$ defined in the normal metallic state. For example, $\mathcal{\check{T}}$  can be observed in  a  thin layer of superconducting proximitized Pt or Ta ($\theta\sim 0.01-0.1$ from Ref.~\onlinecite{sinova2015spin}), which are the typical metals used for measuring the  spin Hall effect. Notice also that the spin Hall angle in superconducting Nb, although small, has been recently quantified in the experiment of  Ref~.\onlinecite{robinson2018}

\section{Summary and outlook} \label{sec:summary}
We have systematically extended the Usadel equation to incorporate  spin-orbit coupling disorder  in diffusive superconducting systems. 
In addition to the spin Hall effect and the swap-current effect describe by the current operator $\check{\cal{J}}_k$, we identified a non-linear spin-orbit coupling induced torque $\check{\cal{T}}$  that has no counterpart in the normal  metallic state.  Interestingly,  the torque is parameterized by  the same  spin Hall angle $\theta$ and   swap current coefficient $\kappa$ that one would obtain in the normal metallic state.
Note that our generalization of Usadel equation accounts for spatially varying kinetic coefficients. By imposing suitable boundary conditions on the Usadel equation, it can readily describe both equilibrium and out-of equilibrium phenomena related to  spin-charge coupling in diffusive hybrid-structures made by superconductor, ferromagnetic and normal metal.



\section*{Acknowledgements:}
We acknowledge useful discussion with Miguel A.~Cazalilla. C.H acknowledges funding by Ministry of Science and Technology, Taiwan (Project No.~107-2917-I-564-009)
 C. H and F.S.B acknowledge funding by the Spanish Ministerio de Econom\'ia y Competitividad (MINECO) (Projects No.~FIS2014-55987-P and FIS2017-82804-P). I.V.T. acknowledges the support by Spanish Ministerio de Economia y Competitividad (MINECO) (Project No. FIS2016-79464-P) and
by the ‘Grupos Consolidados UPV/EHU del Gobierno Vasco’ (Grant No. IT578-13).

\appendix


\section{Derivation of spin-charge coupling self-energy $\hat{\Sigma}^{1a}$ }
 \label{app:sigma}

In this appendix, we detailed the derivation of the self-energy $\check{\Sigma}^{1a}$, as shown in Fig.~3a of the main text.  It can be diagrammatically read off from Fig.~3a  as follow,
\begin{align} 
\check{\Sigma}^{1a}( \mathbf{r_1}, \mathbf{r_2}) 
=& \langle t_{so}(\vec{r}_1)\check{G}(\vec{r}_1,\vec{r}_2) \check{T}^{(0)}(\mathbf{r_2}) \rangle_{\mathrm{imp}} \nonumber \\
+& \langle \check{T}^{(0)}( \mathbf{r_1})\check{G}(\vec{r}_1,\vec{r}_2) t_{so}^{\dagger}(\vec{r}_2) \rangle_{\mathrm{imp}}.
\end{align}
Here  $t_{so}(\vec{r})=-i \lambda^2 \epsilon_{ajk}\sigma_{a} \partial_{j} t_0(\mathbf{r})  \partial_{k} $ is the disorder SOC induced by the impurity potential $t_0(\mathbf{r})$. For short-range randomly distributed impurities, the correlation between the (scalar) disorder potential $t_0(\mathbf{r}_1)$ and the matrix-valued T-matrix, $\check{T}^{(0)}(\vec{r}_2)$ is given by the following: $\langle t_{0}(\mathbf{r_{2}})
 \check{T}^{(0)}(\vec{r}_1) \rangle_{\mathrm{imp}}=\langle t_{0}(\mathbf{r_{1}})
 \check{T}^{(0)}(\vec{r}_2) \rangle_{\mathrm{imp}} 
 =n_{\mathrm{im}}  t_{0}(\mathbf{r_{2}})
 \check{T}^{(0)}(\vec{r}_1) \delta(\vec{r}_1-\vec{r}_2).$
Substitute the impurity average into Eq.~(A1), we arrive at the following:

\begin{widetext}

\begin{equation}
\check{\Sigma}^{1a}(\mathbf{r_{1}},\mathbf{r_{2}})=-in_{\mathrm{im}}\lambda^{2}\epsilon_{ajk}\bigg\{\sigma^{a}\partial_{1k}\check{G}(\mathbf{r_{1}},\mathbf{r_{2}})\partial_{1j}\big[t_{0}(\mathbf{r_{2}})\delta(\mathbf{r_{1}}-\mathbf{r_{2}})\check{T}^{(0)}(\mathbf{r_{2}})\big]-\partial_{2k}\check{G}(\mathbf{r_{1}},\mathbf{r_{2}})\partial_{2j}\big[t_{0}(\mathbf{r_{1}})\delta(\mathbf{r_{1}}-\mathbf{r_{2}})\check{T}^{(0)}(\mathbf{r_{1}})\big]\sigma^{a}\bigg\}
\end{equation}

where the T-matrix (without SOC) is given in Eq.~\eqref{eq:fullT0}:
\begin{equation}
\check{T}^{(0)}( \mathbf{r_1})=  \mathrm{Re }\,t_F(\vec{r}_1) + i \pi N_F \mathrm{Im }\,t_F(\vec{r}_1)  \check{g} ( \mathbf{r_1})
\label{eq:T-matrix0}
\end{equation}
Next, we substitute Eq.~(\ref{eq:T-matrix0}) into the self-energy to arrive at the following equation:
\begin{equation}
\check{\Sigma}^{1a}(\mathbf{r_{1}},\mathbf{r_{2}})=\mathbb{\check{\Sigma}}^{o}(\mathbf{r_{1}},\mathbf{r_{2}})+\mathbb{\check{\Sigma}}^{e}(\mathbf{r_{1}},\mathbf{r_{2}})
\end{equation}
\begin{align}
\mathbb{\check{\Sigma}}^{o}(\mathbf{r_{1}},\mathbf{r_{2}})&=\frac{-i\epsilon_{ajk}}{2\pi N_{F}p_{F}^{2}}\partial_{1j}\delta^{(3)}(\mathbf{r_{1}}-\mathbf{r_{2}})\bigg[\omega_{1}(\mathbf{r_{2}})\sigma^{a}\partial_{1k}\check{G}(\mathbf{r_{1}},\mathbf{r_{2}})+\omega_{1}(\mathbf{r_{1}})\partial_{2k}\check{G}(\mathbf{r_{1}},\mathbf{r_{2}})\sigma^{a}\bigg] \\
\mathbb{\check{\Sigma}}^{e}(\mathbf{r_{1}},\mathbf{r_{2}}) &=\frac{-\epsilon_{ajk}}{2\pi N_{F}  p_{F}^{2}}\partial_{1j}\delta^{(3)}(\mathbf{r_{1}}-\mathbf{r_{2}})\,\bigg[\sigma^{a}\omega_{2}(\mathbf{r_{2}})\partial_{1k}\check{G}(\mathbf{r_{1}},\mathbf{r_{2}})\check{g}(\mathbf{r_{2}})+\omega_{2}(\mathbf{r_{1}})
\partial_{2k}\check{G}(\mathbf{r_{1}},\mathbf{r_{2}})\check{g}(\mathbf{r_{1}})\sigma^{a}\bigg]
\end{align}
Here the derivative always acts on its immediate neighbour.
$\mathbb{\check{\Sigma}}^{o}(\mathbf{r_{1}},\mathbf{r_{2}})$  and $\mathbb{\check{\Sigma}}^{e}(\mathbf{r_{1}},\mathbf{r_{2}})$ are characterized by two scattering rates on the Fermi level: $\omega_1 ( \mathbf{r}_1)= 2\pi n_{\mathrm{im}} N_F \mathrm{Re} \big[ t_{F}^*( \mathbf{r}_1) B( \mathbf{r}_1)\big]$ and  $\omega_2  ( \mathbf{r}_1)= 2\pi n_{\mathrm{im}} N_F \mathrm{Im} \big[ t_{F}^*( \mathbf{r}_1) B( \mathbf{r}_1)\big]$ where  $B( \mathbf{r}_1)=\lambda^2 p_F^2 t_0( \mathbf{r}_1)$ is the SOC scattering vertex. In deriving the above, we used $ \mathrm{Re }\,t_F(\vec{r}_1)= \mathrm{Re }\,t_F^*(\vec{r}_1)$ and $ \mathrm{Im }\,t_F(\vec{r}_1)= -\mathrm{Im }\,t_F^*(\vec{r}_1)$. To proceed
further, we shift all quantities to the center of mass coordinate
$\vec{r}=(\vec{r}_{1}+\vec{r}_{2})/2$ and relative coordinate $\vec{s}=\vec{r}_{1}-\vec{r}_{2}$:

\begin{equation}
\omega_{1}(\mathbf{r_{1}})=\omega_{1}(\mathbf{r})+\frac{s_{i}}{2}\partial_{i}\omega_{1}(\mathbf{r})\;;\;\omega_{1}(\mathbf{r_{2}})=\omega_{1}(\mathbf{r})-\frac{s_{i}}{2}\partial_{i}\omega_{1}(\mathbf{r})
\end{equation}
\begin{equation}
\omega_{2}(\mathbf{r_{1}})=\omega_{2}(\mathbf{r})+\frac{s_{i}}{2}\partial_{i}\omega_{2}(\mathbf{r})\;;\;\omega_{2}(\mathbf{r_{2}})=\omega_{2}(\mathbf{r})-\frac{s_{i}}{2}\partial_{i}\omega_{2}(\mathbf{r})
\end{equation}
\begin{equation}
\partial_{1k}=\frac{1}{2}\partial_{k}+\partial_{s_{k}}\;;\;\partial_{2k}=\frac{1}{2}\partial_k -\partial_{s_{k}}\;;\;\partial_{1j}\delta^{(3)}(\mathbf{r_{1}}-\mathbf{r_{2}})=\partial_{s_{j}}\delta^{(3)}(\vec{s})\;;
\end{equation}
Here $\vec{r}_{\pm}=\vec{r}\pm\vec{s}/2$ and $\partial_k$ ( $\partial_{s_k}$) are the spatial derivative along direction $k$ on the variable $\vec{r}$ ($\vec{s}$).
Using these equations, the self-energy can be written down as follow:
\begin{align}
\mathbb{\check{\Sigma}}^{o}\left(\mathbf{r}_{+} ,\mathbf{r}_{-}\right) & =\frac{-i\epsilon_{ajk}}{2\pi N_{F}p_{F}^{2}}\partial_{s_{j}}\delta^{(3)}(\mathbf{s})\bigg[\left(\omega_{1}(\mathbf{r})-\frac{s_{i}}{2}\partial_{i}\omega_{1}(\mathbf{r})\right)\sigma^{a}\left(\frac{1}{2}\partial_{k}+\partial_{s_{k}}\right)\check{G}\left(\mathbf{r}_{+} ,\mathbf{r}_{-}\right)\nonumber \\
 & +\left(\omega_{1}(\mathbf{r})+\frac{s_{i}}{2}\partial_{i}\omega_{1}(\mathbf{r})\right)\left(\frac{1}{2}\partial_{k}-\partial_{s_{k}}\right)\check{G}\left(\mathbf{r}_{+} ,\mathbf{r}_{-}\right)\sigma^{a}\bigg]\label{eq:app-sigma2b}
\end{align}
\begin{align}
\mathbb{\check{\Sigma}}^{e} \left(\mathbf{r}_{+} ,\mathbf{r}_{-}\right) & =\frac{-\epsilon_{ajk}}{2\pi N_{F}p_{F}^{2}}\partial_{s_{j}}\delta^{(3)}(\mathbf{s})\bigg[\sigma^{a}\left(\omega_{2}(\mathbf{r})-\frac{s_{i}}{2}\partial_{i}\omega_{2}(\mathbf{r})\right)\left(\frac{1}{2}\partial_{k}+\partial_{s_{k}}\right)\check{G}\left(\mathbf{r}_{+} ,\mathbf{r}_{-}\right)\check{g}\left(\mathbf{r}_{-}  \right)\nonumber \\
 & +\left(\omega_{2}(\mathbf{r})-\frac{s_{i}}{2}\partial_{i}\omega_{2}(\mathbf{r})\right)\check{g}\left(\mathbf{r}_{+}  \right)\left(\frac{1}{2}\partial_{k}-\partial_{s_{k}}\right)\check{G}\left(\mathbf{r}_{+} ,\mathbf{r}_{-}\right)\sigma^{a}\bigg]\label{eq:app-sigma1b}
\end{align}
Next, we  Fourier transform the relative coordinate of the self-energy ($\vec{s}$) into momentum $\vec{p}$:
\begin{equation}
\mathbb{\check{\Sigma}}^{o}(\mathbf{r},\vec{p})=\int d\vec{s}\,e^{-i\vec{p}\cdot\vec{s}}\,\mathbb{\check{\Sigma}}^{o} \left(\mathbf{r}_{+} ,\mathbf{r}_{-}\right) \label{eq:app-FT}
\end{equation}
In order to do so, we express all the Green functions in the Wigner coordinates:
\begin{align}
\check{G}\left(\mathbf{r}_{+} ,\mathbf{r}_{-}\right)=\sum_{p'}e^{i\vec{p'}\cdot\vec{s}}\,\check{G}(\mathbf{r},\vec{p}') \label{eq:temp-a} \;  , \;
\check{g}\left(\mathbf{r}_{\pm}\right)=\left( 1\pm \frac{1}{2 }s_i\partial_{i} \right) \,\check{g}\left(\mathbf{r}\right)  
\end{align}

Let us begin by evaluating $\mathbb{\check{\Sigma}}^{o}(\mathbf{r},\vec{p})$.
We substitute the first equation in Eq.~(\ref{eq:temp-a}) into Eq.~(\ref{eq:app-sigma2b})
and apply the Fourier transform in Eq.~(\ref{eq:app-FT}) to arrive
at the following:
\begin{equation}
\mathbb{\check{\Sigma}}^{o}(\mathbf{r},\vec{p})=\frac{-i\epsilon_{ajk}}{2\pi N_{F}p_{F}^{2}}\sum_{\vec{p}'}\,i(p_{j}-p'_{j})\omega_{1}(\mathbf{r})\left(\frac{1}{2}\big\{\sigma^{a},\partial_{k}\check{G}\left(\mathbf{r},\vec{p}'\right)\big\}+i\big[\sigma^{a},p_{k}^{'}\check{G}\left(\mathbf{r},\vec{p}'\right)\big]\right)+\frac{i\partial_{j}\omega_{1}(\mathbf{r})}{2}\big\{\sigma^{a},p_{k}^{'}\check{G}\left(\mathbf{r},\vec{p}'\right)\big\}
\end{equation}
The intricate momentum summation above can be done in the quasiclassical
limit. For $p_{F}^{-1}$ to be much smaller than the typical variation of spectral weights and observables, the Green functions will be peaked at the Fermi level so their momentum summation can be simplify as follow:
\begin{equation}
\sum_{\vec{p}'}\check{G}\left(\mathbf{r},\vec{p}\right)=-i\pi N_{F} \check{g}\left(\mathbf{r}\right)
\;;\;\sum_{\vec{p}'}\check{G}\left(\mathbf{r},\vec{p}'\right)p_{k}'=-i\pi N_{F}\frac{p_{F}}{3} \check{g}_k (\mathbf{r})
\label{eq:app-anti}
\end{equation}
Here $\check{g}\left(\mathbf{r}\right)$ is the isotropic (zeroth-moment) of the Green function while   $\check{g}_k\left(\mathbf{r}\right)$  is the  (first-moment) of the Green function along spatial direction $k$. Since
the relevant self-energy that enters the collision integral has its
momentum parked at the Fermi level, we set $\mathbb{\check{\Sigma}}^{o}(\mathbf{r},\vec{p}=p_{F}\vec{n})=\mathbb{\check{\Sigma}}^{o}(\mathbf{r},\vec{n})$
and arrive at the equation quoted in the main text:
\begin{align}
\mathbb{\check{\Sigma}}^{o}(\vec{n},\mathbf{r})= & \frac{\omega_{1}(\mathbf{r})\epsilon_{ajk}}{2}\bigg(n_{k}\frac{1}{3}\left[\check{g}_{j}(\mathbf{r}),\sigma^{a}\right]+\frac{1}{2p_{F}}n_{k}\left\{ i\partial_{j}\check{g}(\mathbf{r}),\sigma^{a}\right\} +\frac{1}{6p_{F}}i\partial_{k}\left\{ \check{g}_{j}(\mathbf{r}),\sigma^{a}\right\} \bigg)  -\frac{i\epsilon_{ajk}\partial_{j}\omega_{1}(\mathbf{r})}{12p_{F}}\{\sigma^{a},\check{g}_{k}(\mathbf{r})\}
\end{align}
Note that $\mathbb{\check{\Sigma}}^{o}(\mathbf{r},\vec{p})$ is now
an algebraic equation expressed in terms of the zeorth and first moment
of the quasiclassical Green function, this is a tremendous simplification compare to the previous equation where $\mathbb{\check{\Sigma}}^{o}(\mathbf{r},\vec{p})$ is a \emph{functional} of the Green function. 
$\mathbb{\check{\Sigma}}^{e}(\mathbf{r},\vec{p})$ can be derived following the same procedure: we substitute Eqs.~(\ref{eq:temp-a}) and 
into Eq.~(\ref{eq:app-sigma1b}) and apply the Fourier transform Eq.~(\ref{eq:app-FT})
to arrive at the following:
\begin{align}
\mathbb{\check{\Sigma}}^{e}(\vec{p},\mathbf{r})= \,& \frac{-\epsilon_{ajk}\omega_{2}(\mathbf{r})}{2\pi N_{F} p_{F}^{2}}\sum_{\vec{p}'}\,i(p_{j}-p'_{j})\left(\sigma^{a}\left(\frac{\partial_{k}}{2}+ip'_{k}\right)\check{G}\left(\mathbf{r},\vec{p}'\right)\check{g}\left(\mathbf{r}\right)+\check{g}\left(\mathbf{r}\right)\left(\frac{\partial_{k}}{2}-ip'_{k}\right)\check{G}\left(\mathbf{r},\vec{p}'\right)\sigma^{a}\right)\nonumber \\
+ & \frac{-\epsilon_{ajk}\omega_{2}(\mathbf{r})}{  2\pi N_{F} p_{F}^{2} }\sum_{\vec{p}'}\,\frac{1}{2}\left(\sigma^{a}\left(\frac{\partial_{k}}{2}+ip'_{k}\right)\check{G}\left(\mathbf{r},\vec{p}'\right)\partial_{j}\check{g}\left(\mathbf{r} \right)-\partial_{j}\check{g}\left(\mathbf{r}\right)\left(\frac{\partial_{k}}{2}-ip'_{k}\right)\check{G}\left(\mathbf{r},\vec{p}'\right)\sigma^{a}\right)\nonumber \\
+ & \frac{-\epsilon_{ajk}}{ 2\pi N_{F} p_{F}^{2} }\sum_{\vec{p}'}\,\frac{\partial_{j}\omega_{2}(\vec{r})}{2}ip_{k}^{'}\left(\sigma^{a}\check{G}\left(\mathbf{r},\vec{p}'\right)\check{g}\left(\mathbf{r}\right)+\check{g}\left(\mathbf{r} \right) \check{G}\left(\mathbf{r},\vec{p}'\right)\right)
\end{align}
Next, we use  Eq.~(\ref{eq:app-anti}) to perform the momentum integration and set $\vec{p}=p_F \vec{n}$ to arrive at the result in the main text.

\section{Collision integral of the Usadel Equation} \label{app:coll-int}

In this appendix, we explain in detail how we arrive at the result $\mathcal{I}_{0}^{(1)}=\check{\mathcal{T}}-\partial_{k}\mathcal{\check{J}}_{k}^{\text{an}}$ shown in Eq.~\eqref{eq:coll-int-g0},
from the self-energy. The collision integral in the Usadel equation (the zeroth moment of the kinetic equation) is defined by Eq.~(\ref{eq:coll-inta}) of the main text:
\begin{equation}
\check{\mathcal{I}}_{0}[\check{g},\check{g}_{k}]=-i\left\langle \left[\check{\Sigma},\,\check{\mathrm{g}} (\mathbf{n},\mathbf{r})\right]\right\rangle -\frac{\partial_{k}}{2}\left\langle\big\{\partial_{p_{k}}\check{\Sigma}\big|_{p=p_{F}}\,,\,\check{\mathrm{g}} (\mathbf{n},\mathbf{r}) \big\}\right\rangle.
\label{eq:app-coll-int-def}
\end{equation}
Note $\check{\mathrm{g}}=\check{\mathrm{g}}  (\mathbf{n},\mathbf{r})$ is the Eilenberger GF and readers should not be confused with the zeroth moment (Usadel) GF $\check{g}=\check{g} (\mathbf{r})$.
To  linear order in SOC, there are two class of self-energy diagrams shown in Fig.~3 of the main text which are denoted as $\check{\Sigma}^{1a}$ and $\check{\Sigma}^{1b}$. The corresponding contribution to $\check{\mathcal{I}}_{0}$ are calculated below.

\subsection{Contribution of $\check{\Sigma}^{1a}$ to the collision integral $\mathcal{I}_{0}^{(1)}$}

We first consider the self-energy shown in Fig.~3a which we defined as $\check{\Sigma}^{1a}$. From Appendix.~\ref{app:sigma} and the main text, $\check{\Sigma}^{1a}=\mathbb{\check{\Sigma}}^{o}+\mathbb{\check{\Sigma}}^{e}$ reads as follows
\begin{align}
\mathbb{\check{\Sigma}}^{o}(\vec{n},\mathbf{r})= & \frac{\omega_{1}\epsilon_{ajk}}{2}\bigg(\frac{1}{3}n_{k}\left[\check{g}_{j},\sigma^{a}\right]+\frac{1}{2p_{F}}n_{k}\left\{ i\partial_{j}\check{g},\sigma^{a}\right\} +\frac{1}{6p_{F}}i\partial_{k}\left\{ \check{g}_{j},\sigma^{a}\right\} \bigg)-\frac{i\epsilon_{ajk}\partial_{j}\omega_{1}}{12p_{F}}\{\sigma^{a},\check{g}_{k}\}\label{eq:app-sigma1}\\
\mathbb{\check{\Sigma}}^{e}(\vec{n},\mathbf{r})= & \frac{-i\epsilon_{ajk}\omega_{2}}{2}\bigg(\frac{1}{3}n_{j}\left(\sigma^{a}\check{g}_{k}\check{g}-\check{g}\check{g}_{k}\sigma^{a}\right)-\frac{n_{j}}{2p_{F}}\left(\sigma^{a}(i\partial_{k}\check{g})\,\check{g}+\check{g}\,(i\partial_{k}\check{g})\sigma^{a}\right)+\frac{i\partial_{k}}{6p_{F}}\left(\sigma^{a}\check{g}_{j}\,\check{g}+\check{g}\,\check{g}_{j}\sigma^{a}\right)\bigg)\nonumber \\
 & -\frac{\epsilon_{ajk}\partial_{j}\omega_{2}}{12p_{F}}\bigg(\sigma^{a}\,\check{g}_{k}\,\check{g}+\check{g}\,\check{g}_{k}\,\sigma^{a}\bigg)\label{eq:app-sigma22}
\end{align}
In the following, we substitute each of the above contributions into the collision integral of the
Usadel equation:
\begin{equation}
\check{\mathcal{I}}_{0}^{(1a)}[\check{g},\check{g}_{k}]=-i\left\langle \left[\check{\Sigma}^{1a},\,\check{\mathrm{g}}\right]\right\rangle -\frac{\partial_{k}}{2}\left\langle \big\{\partial_{p_{k}}\check{\Sigma}^{1a}\big|_{p=p_{F}}\,,\,\check{\mathrm{g}}\big\}\right\rangle=\check{\mathcal{I}}^{\text{(e)}}+\check{\mathcal{I}}^{\text{(o)}}
\label{eq:app-coll-int}
\end{equation}
Let us begin with $\mathbb{\check{\Sigma}}^{o}(\vec{n},\mathbf{r})$.
We substitute Eq.~\eqref{eq:app-sigma1} into Eq.~\eqref{eq:app-coll-int}
and arrive at the following result:
\begin{align}
\check{\mathcal{I}}^{\text{(o)}}= & \frac{i\omega_{1}\epsilon_{ajk}}{18}\big[\sigma^{a},\check{g}_{j}\,\check{g}_{k}\big]+\frac{\omega_{1}\epsilon_{ajk}}{12p_{F}}\bigg(-\big[\{\partial_{k}\check{g},\sigma^{a}\}\,,\,\check{g}_{j}\big]+\big[\{\partial_{k}\check{g}_{j},\sigma^{a}\}\,,\,\check{g}\big]-\big\{[\partial_{k}\check{g}_{j},\sigma^{a}]\,,\,\check{g}\big\}-\big\{[\check{g}_{j},\sigma^{a}]\,,\,\partial_{k}\check{g}\big\}\bigg)\nonumber \\
 & -\frac{(\partial_{j}\omega_{1})\,\epsilon_{ajk}}{6p_{F}}\left(\sigma^{a}\check{g}_{k}\,\check{g}-\check{g}\,\check{g}_{k}\sigma^{a}\right).\label{eq:app-I-2b}
\end{align}
The collision integral of the Usadel equation in Ref.~\onlinecite{bergeret2016}
is obtained by considering only the first line above. However, there
is typo in the supplementary material of  Ref.~\onlinecite{bergeret2016} where we
skipped the first term in $\check{\mathcal{I}}^{\text{(o)}}$. We
go beyond  Ref.~\onlinecite{bergeret2016} and include spatial variation of $\omega_{1}$.
Note that the last two term in the first line of Eq.~\eqref{eq:app-I-2b}  arised from the anticommutator in Eq.~\eqref{eq:app-coll-int}. They contribute to the same order of magnitude as the third and fourth term  in the first line of Eq.~\eqref{eq:app-I-2b}. Therefore, as emphasized in the main text, when the system is subjected to disorder SOC, it is crucial to retain the Poisson bracket in the collision integral of the kinetic equation.

Recall that the constitutive relation obtained in the main text is
given by the following:
\begin{equation}
\check{g}_{k}=-l\check{g}\partial_{k}\check{g}+\frac{l}{2}\epsilon_{akj}\big(\theta'\left\{ \partial_{j}\check{g},\sigma^{a}\right\} +i\kappa'\left[\check{g}\partial_{j}\check{g},\sigma^{a}\right]\big)\label{eq:app-consti}
\end{equation}
To proceed further, we substitute Eq.~(\ref{eq:app-consti}) into (\ref{eq:app-I-2b}).
To linear in SOC, this amounts to replacing $\check{g}_{k}=-l\check{g}\partial_{k}\check{g}$
in Eq. (\ref{eq:app-I-2b}) and the resulting equation reads:
\begin{equation} \label{eq:app-Io}
\check{\mathcal{I}}^{\text{(o)}}=\frac{1}{2}\epsilon_{akj}\left(\frac{\omega_{1}l}{3p_{F}}\,\big[\sigma^{a},\check{g}\partial_{k}\check{g}\partial_{j}\check{g}\big]+\left(\frac{\omega_{1}l^{2}}{9}\right)i\big[\partial_{k}\check{g}\partial_{j}\check{g},\sigma^{a}]-\partial_{k}\left(\frac{\omega_{1}l}{3p_{F}}\,\big\{\partial_{j}\check{g},\sigma^{a}\big\}\right)\right)
\end{equation}
Importantly, the space derivative (i.e. $\partial_{k}$) in the last
term applies to the quantities inside the bracket as chain-rules,
i.e. it is a total divergence. Recall that $\omega$ and $l$ depend on the spatially coordinate $\vec{r}$. In deriving the above, we used 
the property $\big\{\check{g}\,,\,\partial_{k}\check{g}\big\}=0$
which follows from the normalization condition $\check{g}^{2}=1$ we found in Eq.~\eqref{eq:norm-presv}.
Next, we use $D=v_{F}l/3$ to simplify the coefficients above to arrive
at the following equation:
\begin{equation}
\check{\mathcal{I}}^{\text{(o)}}=\frac{D}{4}\epsilon_{akj}\left(\frac{2\omega_{1}\tau}{p_{F}l}\,\big[\sigma^{a},\check{g}\partial_{k}\check{g}\partial_{j}\check{g}\big]+\left(\frac{2}{3}\omega_{1}\tau\right)i\big[\partial_{k}\check{g}\partial_{j}\check{g},\sigma^{a}]\right)-\partial_{k}\left(\frac{D\omega_{1}\tau}{2p_{F}l}\epsilon_{akj}\,\big\{\partial_{j}\check{g},\sigma^{a}\big\}\right)
\end{equation}
The first two terms correspond to part of the spin-orbit torque $\check{\mathcal{T}}$
while the last term corresponds to the anticommutator part of $\mathcal{\mathcal{\check{J}}^{\text{an}}}_{k}$,
c.f. Eq.~\eqref{eq:coll-int-g0}

Next, we follow the same logic above to derive the collision integral
from Eq.~(\ref{eq:app-sigma22}). First, we substitute Eq.~(\ref{eq:app-sigma22})
into Eq.~(\ref{eq:app-coll-int}) and replace $\check{g}_{k}=-l\check{g}\partial_{k}\check{g}$, the resulting equation reads:
\begin{equation}
\check{\mathcal{I}}^{\text{(e)}}=\epsilon_{ajk}\omega_{2}\left(-\frac{l^{2}}{18}\big[\sigma^{a},\partial_{j}\check{g}\partial_{k}\check{g}\big]-\frac{l^{2}}{18}\partial_{j}\check{g}[\sigma^{a}\,,\check{g}]\partial_{k}\check{g}+\frac{l}{6p_{F}}\,i\big[\sigma^{a},\partial_{j}\check{g}\partial_{k}\check{g}\big]-\frac{i\partial_{k}}{6p_{F}}\left(l\big[\check{g}\partial_{j}\check{g},\sigma^{a}\big]\right)\right)-\frac{\epsilon_{ajk}l}{6p_{F}}i\big[\check{g}\partial_{j}\check{g},\sigma^{a}\big]\partial_{k}\omega_{2}
\end{equation}
In the above, the last two terms can be combined into a total divergence. Next, we use $D=v_{F}l/3$ to simplify the coefficients above and arrive at the following equation:
\begin{equation} \label{eq:app-Ie}
\check{\mathcal{I}}^{\text{(e)}}=\frac{D}{4}\epsilon_{akj} \bigg(
 \frac{2}{3} \omega_2 \tau   
\,\big[\sigma^{a},\check{g} \partial_{k}\check{g} \partial_{j}\check{g} \big] 
  -\frac{\omega_2 \tau}{p_Fl}   
i\big[\partial_{k}\check{g}\partial_{j}\check{g},\sigma^{a}] \bigg)
 -\partial_k \left(  \epsilon_{akj}    \frac{D\omega_2 \tau}{2p_Fl }  \,i \,[\sigma^a, \check{g} \partial_j \check{g} ]  \right) -\epsilon_{ajk} \frac{D \omega_2 \tau }{6}\partial_{j}\check{g}[\sigma^{a}\,,\check{g}]\partial_{k}\check{g}
\end{equation}

Let us now combine Eq.~\eqref{eq:app-Io} and Eq.~\eqref{eq:app-Ie} and arrive at the following:
\begin{equation} \label{eq:app-I-final}
\check{\mathcal{I}}^{\text{(e)}}+\check{\mathcal{I}}^{\text{(o)}}=\check{\mathcal{T}}-\partial_{k}\mathcal{\check{J}}_{k}^{\text{an}}-\epsilon_{ajk} \frac{D \omega_2 \tau }{6}\partial_{j}\check{g}[\sigma^{a}\,,\check{g}]\partial_{k}\check{g}
\end{equation}
The last term stems from the second term in $\check{\mathcal{I}}^{\text{(e)}}$.  As we will show in next section, as a consequence of the optical theorem, this term is canceled exactly by the contribution coming from the self-energy $\check{\Sigma}^{1b}$ shown in Fig.~3b of the main text.

\subsection{Contribution of $\check{\Sigma}^{1b}$ to the collision integral $\mathcal{I}_{0}^{(1)}$}

The expression for $\check{\Sigma}^{1b}$ is derived in the main text and given by Eqs.~(\ref{eq:sigma1b-fin})-(\ref{tildeK}) which we repeat here:
\begin{equation}
\check{\Sigma}^{1b}=-in_{\rm im}\lambda^{2}t_{0}(\pi N_{F}p_{F})^{2}\bigg((\text{Re\,}t_{F})^{2}\tilde{\mathcal{K}}-(\text{Im\,}t_{F})^{2}\check{g}\tilde{\mathcal{K}}\check{g}+(\text{Im\,}t_{F}\text{Re\,}t_{F})\left(\check{g}\tilde{\mathcal{K}}+\tilde{\mathcal{K}}\check{g}\right)\bigg)\label{eq:app-3b-wig}
\end{equation}

\begin{equation}
\tilde{\mathcal{K}}=\epsilon_{ajk}\Big[\frac{1}{9}\check{g}_{j}\sigma^{a}\check{g}_{k}+\frac{i}{6p_{F}}\partial_{j}\check{g}\sigma^{a}\check{g}_{k}-\frac{i}{6p_{F}}\check{g}_{j}\sigma^{a}\partial_{k}\check{g}\Big].
\end{equation}
As stated in the main text, $\check{\Sigma}^{1b}$ does
not carry the external momentum. As a result it does not contribute to
the collision integral $\mathcal{\check{I}}_k$ of the constitutive relation as it would
lead to terms of the order of $\epsilon_{1}^{3}$ and $\epsilon_{1}^{2}\epsilon_{2}$ which are irrelevant in our approximation.
Secondly, when $\check{\Sigma}^{1b}$ is substituted into the collision integral of Eq.~\ref{eq:app-coll-int-def}, the anticommutator part, which is proportional
to $\partial_{p}\Sigma(\vec{p},\vec{r})$, vanishes and we are left with the following simple results
\begin{equation}
\mathcal{\check{I}}^{1b}=-i\big[\check{\Sigma}^{1b}\,,\,\check{g}\big].
\end{equation}
Next, using the normalization condition (i.e.~$\check{g}^{2}=1$) we showed from Eq.~\eqref{eq:norm-presv}, the collision integral can be greatly simplified. First, note that the last term in Eq.~(\ref{eq:app-3b-wig}) does not contribute to $\mathcal{\check{I}}^{1b}$ because
\begin{equation}
\big[\check{g}\tilde{\mathcal{K}}+\tilde{\mathcal{K}}\check{g},\check{g}\big]=\check{g}\tilde{\mathcal{K}}\check{g}+\tilde{\mathcal{K}}\check{g}^{2}-\check{g}^{2}\tilde{\mathcal{K}}-\check{g}\tilde{\mathcal{K}}\check{g}=0.
\end{equation}
Note also that because $\big[\check{g}\tilde{\mathcal{K}}\check{g},\check{g}\big]=-[\tilde{\mathcal{K}},\check{g}]$ the second term in Eq.~\ref{eq:app-3b-wig} reduces to the form similar to the first term. Hence, by collecting these results, the collision integral can be simplified as follows
\begin{equation}
\mathcal{\check{I}}^{1b}=-n_{\rm im}\lambda^{2}t_{0}(\pi N_{F}p_{F})^{2}|t_{F}|^2\big[\tilde{\mathcal{K}}\,,\,\check{g}\big].
\end{equation}
To  linear order in SOC, we substitute $\check{g}_{k}=-l\check{g}\partial_{k}\check{g}$
to evaluate $\big[\tilde{\mathcal{K}}\,,\,\check{g}\big]$ and the
resulting collision integral reads:
\begin{equation}
\mathcal{\check{I}}^{1b}=n_{\rm im}\lambda^{2}\epsilon_{ajk}t_{0}\left(\frac{\pi N_{F}p_{F}l}{3}\right)^{2}|t_{F}|^2\partial_{j}\check{g}[\sigma^{a},\check{g}]\partial_{k}\check{g}
\end{equation}
This has exactly the same matrix structure as the last term in Eq.~\eqref{eq:app-I-final}. 
To see how the coefficients add up, we invoke the optical theorem of Eq.~(\ref{eq:opti-theorem}) which follows from unitarity of the scattering S-matrix:
\begin{equation}
\pi N_{F}|t_{F}|^2=-\text{Im\,}t_{F}.
\end{equation}
This brings our collision integral to the following form
\begin{equation}
\mathcal{\check{I}}^{1b}=\epsilon_{ajk}\frac{l^{2}\omega_2}{18}\partial_{j}\check{g}[\sigma^{a},\check{g}]\partial_{k}\check{g}=\epsilon_{ajk}\frac{D\omega_{2}\tau}{6}\partial_{j}\check{g}[\sigma^{a},\check{g}]\partial_{k}\check{g},
\end{equation}
where we used the definition $\omega_2=2\pi n_{\mathrm{im}}N_{F}\text{Im}t_{F}^{*}\lambda^{2}p_{F}^{2}t_{0}$. 

Finally, by adding the derived $\mathcal{\check{I}}^{1b}$ to the collision integral from $\check{\Sigma}^{1a}=\mathbb{\check{\Sigma}}^{o}+\mathbb{\check{\Sigma}}^{e}$, Eq.~\eqref{eq:app-I-final}, we arrive at the result quoted in Eq.~\eqref{eq:coll-int-g0} of the main text:
\begin{equation}
\check{\mathcal{I}}^{\text{(e)}}+\check{\mathcal{I}}^{\text{(o)}}+\mathcal{\check{I}}^{1b}=\check{\mathcal{T}}-\partial_{k}\mathcal{\check{J}}_{k}^{\text{an}} \equiv \check{\mathcal{I}}^{(1)}_{0}
\end{equation}
This is the full collision integral of the Usadel equation at the linear order in SOC.

\end{widetext}

\bibliography{ref-SC-SOC.bib}
\end{document}